\pdfoutput=1
\documentclass[usenatbib,usedAMS]{mnras}

\usepackage{natbib}
\usepackage{amsmath}
\usepackage{graphicx}
\usepackage{color}
\usepackage{mathrsfs}
\usepackage{epsfig}
\usepackage{comment}
\usepackage{ulem}

\newcommand{\msun}{{\rm M}_\odot}

\newcommand{\cc}{{\rm cm}^{-3}}

\newcommand{\msunyr}{{\rm M}_\odot~{\rm yr}^{-1}}

\newcommand{\kms}{{\rm km~s}^{-1}}
\newcommand{\rB}{R_{\rm B}}
\newcommand{\rC}{R_{\rm C}}
\newcommand{\MMB}{\dot{M}/\dot{M}_{\rm B}}
\newcommand{\MME}{\dot{M}/\dot{M}_{\rm Edd}}
\newcommand{\MBME}{\dot{M}_{\rm B}/\dot{M}_{\rm Edd}}
\newcommand{\mdotE}{\dot{M}_{\rm Edd}}
\newcommand{\mdotB}{\dot{M}_{\rm B}}
\newcommand{\dotM}{\dot{M}}

\newcommand{\K}{{\rm K}}
\newcommand{\beq}{\begin{equation}}
\newcommand{\eeq}{\end{equation}}

\defcitealias{K14}{K14}
\defcitealias{K16}{K16}
\defcitealias{Visbal_2015}{VHB15}

\title[A global solution of rotating accretion gas]
{Low density, radiatively inefficient rotating-accretion flow onto a black hole}

\author[]{Kohei Inayoshi$^{1}$\thanks{E-mail: inayoshi@astro.columbia.edu (KI)}
\thanks{Simons Society of Fellows, Junior Fellow.},
{Jeremiah P. Ostriker}$^1$, {Zolt\'an Haiman}$^1$ and Rolf Kuiper$^2$
\\
$^1$Department of Astronomy, Columbia University, 550 W. 120th Street, New York, NY 10027, USA,\\
$^2$Institute of Astronomy and Astrophysics, University of T{\"u}bingen, Auf der Morgenstelle 10, D-72076 T{\"u}bingen, Germany
}

\begin{document}
\maketitle
\label{firstpage}

\begin{abstract}
We study low-density axisymmetric accretion flows onto black holes (BHs) 
with two-dimensional hydrodynamical simulations, adopting 
the $\alpha$-viscosity prescription.
When the gas angular momentum is low enough to form a rotationally supported disk
within the Bondi radius ($\rB$), we find a global steady accretion solution.
The solution consists of a rotational equilibrium distribution around $r\sim \rB$,
where the density follows $\rho \propto (1+\rB/r)^{3/2}$,
surrounding a geometrically thick and optically thin accretion disk at the centrifugal radius $\rC(<\rB)$, 
where thermal energy generated by viscosity is transported via convection.
Physical properties of the inner solution agree with those expected in convection-dominated accretion flows 
(CDAF; $\rho \propto r^{-1/2}$).
In the inner solution, the gas inflow rate decreases towards the center due to convection ($\dotM \propto r$),
and the net accretion rate (including both inflows and outflows) is strongly suppressed
by several orders of magnitude from the Bondi accretion rate $\mdotB$.
The net accretion rate depends on the viscous strength, following 
$\MMB \propto (\alpha/0.01)^{0.6}$.
This solution holds for low accretion rates of $\MBME \la 10^{-3}$
having minimal radiation cooling, where $\mdotE$ is the Eddington accretion rate.
In a hot plasma at the bottom
($r<10^{-3}~\rB$), 
thermal conduction would dominate the convective energy flux.
Since suppression of the accretion by convection ceases, 
the final BH feeding rate is found to be $\MMB \sim 10^{-3}-10^{-2}$.
This rate is as low as $\MME \sim 10^{-7}-10^{-6}$ 
inferred for SgrA$^*$ and the nuclear BHs in M31 and M87, and 
can explain their low luminosities, without invoking any feedback mechanism.
\end{abstract}



\section{Introduction}
\label{sec:intro}

The basic physics of gas accretion onto an astrophysical object was first studied 
for a spherically symmetric flow \citep{Bondi_1952}.
When rotation is not included, gas accretion onto a black hole (BH) with a mass of $M_\bullet$
begins from a characteristic radius,
where the negative gravitational energy becomes greater than the thermal energy of the gas.
The so-called Bondi radius is given by
\begin{equation}
R_{\rm B}\equiv \frac{GM_\bullet}{c_{\rm s}^2},
\end{equation}
where $c_{\rm s}$ is the sound speed of the gas.
One can safely assume roughly free fall within this radius at a rate of $\dot{M}_{\rm B}$,
so-called the Bondi rate. 
If radiation emitted from the inner region of the flow exceeds 
several percent of the Eddington luminosity 
($L_{\rm Edd}\equiv 4\pi cGM_\bullet / \kappa_{\rm es}$, 
where $\kappa_{\rm es}$ is the electron scattering opacity, then this “feedback” can add enough
energy and momentum at the Bondi radius to reverse the flow 
\citep{Ostriker_1976,Ciotti_Ostriker_2001,Proga_2007,
Ciotti_2009,Milosavljevic_2009,PR_2011,IHO_2016}.
However, the problem is rather simple for low accretion rates \citep{Shapiro_1973,Park_1990a}.

Now let us add angular momentum per unit mass $j$ to the flow. 
This effect introduces a new physical scale of
\begin{equation}
R_{\rm C}\equiv \frac{j^2}{GM_\bullet},
\end{equation}
which is the centrifugal radius where the centrifugal force balances the gravity of the BH. 
If this radius is larger than the Schwarzschild radius
\begin{equation}
R_{\rm Sch}\equiv \frac{GM_\bullet}{c^2},
\end{equation}
the accreting gas will simply settle into a rotational equilibrium distribution with no net inflow
\citep{Fishbone_Moncrief_1976,Papaloizou_Pringle_1984,LOS_2013}.
However, if viscosity exists, then angular momentum transport is possible and 
an accretion disk forms at the bottom of the distribution.
The effective shear viscosity driven by magneto rotational instability (MRI) is often described with
the standard $\alpha$-prescription \citep[e.g.,][]{SS_1973} as 
\begin{equation}
\nu =\alpha c_{\rm s}H, 
\end{equation}
where $H$ is the scale-height of the inner flow and $\alpha$ presents the strength of viscosity
estimated as $\alpha \sim O(10^{-2})$ by magneto-hydrodynamical (MHD) simulations 
\citep{Balbus_Hawley_1991,Matsumoto_1995,Stone_1996,Balbus_Hawley_1998,Sano_2004}.

There have been a number of analytical solutions and numerical studies 
for rotating flows with viscous angular momentum transport (e.g., \citealt{SS_1973}; 
see reviews by \citealt{Pringle_1981,Kato_2008}, references therein). 
Among those, we here focus on accretion flows which cannot lose internal energy 
via radiative cooling because of very low gas density.
Such radiatively inefficient accretion flows are quite interesting since 
many observed BHs accrete at rates of only a small fraction of the Bondi accretion rate
and their radiation luminosity is as low as $\sim 10^{-1}-10^{-9}~L_{\rm Edd}$
\citep{Ho_2008,Ho_2009}.
Sagittarius A$^*$ (Sgr A$^*$) is inferred to be accreting at a rate of 
$10^{-3}$ to $10^{-2}~\dot{M}_{\rm B}$ \citep[e.g.,][]{Yuan_2003,Quataert_2004},
where $\dot{M}_{\rm B}\simeq 10^{-5}~\msunyr $
is measured from the temperature and density near the Bondi radius with 
X-ray observations \citep{Baganoff_2003}.
Because of such a low accretion rate, the bolometric luminosity of Sgr A$^*$ 
($M_\bullet \simeq 4\times 10^6~\msun$; \citealt{Ghez_2003}) is as small as 
$L_{\rm bol}\sim 10^{36}~{\rm erg~s}^{-1}\sim 2\times 10^{-9}~L_{\rm Edd}$.
The second example is a BH at the center of the giant elliptical galaxy M87.
The gas accretion rate at the vicinity of the BH is estimated as
$\la 9.2\times 10^{-4}~\msunyr$ \citep{Kuo_2014}, which is lower than 
$\sim 10^{-2}~\dot{M}_{\rm B}$ \citep{Russell_2015}.
Since the BH mass is estimated as $M_\bullet = 6.6^{+0.4}_{-0.4}\times 10^9~\msun$ 
\citep{Gebhardt_2011} and $M_\bullet = 3.5^{0.9}_{-0.7}\times 10^9~\msun$ \citep{Walsh_2013},
the bolometric luminosity of $L_{\rm bol} \sim 2\times 10^{41}$ erg s$^{-1}$ is 
$\sim 3\times 10^{-7}~L_{\rm Edd}$.
The third example is a BH at the center of the Andromeda Galaxy (M31).
The estimated BH mass is $M_\bullet \simeq 1.4^{+0.7}_{-0.3}\times 10^8~\msun$ \citep{Bender_2005}.
The Bondi accretion rate and the X-ray luminosity are estimated as $\dot{M}_{\rm B}\simeq 7\times 10^{-5}~\msunyr$
and $L_{\rm X}\simeq 2\times 10^{36}$ erg s$^{-1}\simeq 10^{-10}~L_{\rm Edd}$, respectively \citep{Garcia_2010}.
The corresponding bolometric luminosity is inferred as $\simeq 10^{-9}~L_{\rm Edd}$ by
assuming the bolometric correction factor to be $\simeq 10$ \citep{Hopkins_2007}.

In very low-accretion-rate flows, the gas never cools via radiation and 
the accretion disk becomes hot and thick.
This thick-disk solution has been found by \cite{Ichimaru_1977} and 
studied in the subsequent works by \cite{NY_1994,NY_1995}.
This solution is well-known in a simplified self-similar version 
as advection-dominated accretion flows (ADAFs).
Their physical properties can be understood by considering a self-similar solution,
which suggests that low-density accretion flow is hot and geometrically extended 
in the polar direction and the thermal energy is advected with the flow onto the BH.
Moreover, as a related solution, the so-called adiabatic inflow-outflow solution (ADIOS) 
has been proposed by \cite{Blandford_Begelman_1999,Blandford_Begelman_2004}, 
where matter can accrete at very low rates due to strong outflows which carry energy 
and angular momentum away.

As pointed out by  \cite{NY_1994,NY_1995}, ADAF solutions tend to be convectively unstable
because the gas entropy increases towards the center due to thermal energy release via viscosity.
\cite{Narayan_2000} and \cite{Quataert_2000} conducted stability analyses 
for rotating adiabatic gas against convection motions, and found that
the density profile in a marginally stable state follows $\rho \propto r^{-1/2}$, 
whose slope is rather flatter than $\rho \propto r^{-3/2}$ expected from ADAFs.
This solution is known as convection-dominated accretion flows (CDAF).
Numerical simulations of radiatively inefficient accretion flows have also supported
that CDAF solutions can exist in a weak viscosity and no cooling regime
\citep{IA_1999,IA_2000,Stone_1999,INA_2003}.
Many previous works with multi-dimensional MHD simulations have studied 
the properties of the accretion flow at the vicinity of the central BH 
($r\la {\rm a~few}\times 10^2~R_{\rm Sch}$)
starting from torus-like initial reservoirs of gas
(\citealt{Stone_Pringle_2001,Hawley_2001,Machida_2001,McKinney_&_Gammie_2004,
Ohsuga_2009,INA_2003, Narayan_2012}; see also \citealt{Yuan_2012a}, who studied 
the gas dynamics over a wider range of spacial scales).
However, because of the limitation of the computational domain,
it remains unclear how and whether the accretion flow, which begins from large radii ($r\sim R_{\rm B}$),
can lose the angular momentum and reach smaller radii ($r\ll R_{\rm B}$), 
where the gas is tightly bound to the central BH.

To explicitly connect large and small scales in a self-consistent solution, we study dynamics of gas accretion onto a BH 
over a wide range of spatial scales, and are able to connect the inner region, 
where a disk forms, with the region well outside the Bondi radius, 
where the accretion originates in the first place.
\cite{LOS_2013} studied a similar problem for one specific parameter set,
i.e., $R_{\rm C}/R_{\rm B}=0.02$ and $\alpha \sim 10^{-3}$,
with two dimensional hydrodynamical simulations assuming equatorial symmetry.
This work followed gas accretion from outside the Bondi radius to within the centrifugal radius.
\cite{Proga_2003b} have conducted two-dimensional MHD simulations for this problem
with a more limited dynamic range ($r<10^3~R_{\rm Sch}$), 
and found that gas accretion 
is allowed by viscosity driven by the MRI.
However, several crucial questions remain unanswered.
{\it How do physical parameters, e.g., $R_{\rm C}/R_{\rm B}$ and $\alpha$, 
determine the actual inflow rate onto a BH?
What do physical properties of the accretion flow inside the centrifugal radius look like?}

We perform two-dimensional viscous-hydrodynamical simulations with a large dynamic range
including angular momentum transport with the $\alpha$-viscosity prescription,
and address the above questions for radiatively inefficient rotating accretion flows onto a BH. 
Here, we focus on cases where the gas angular momentum is low enough 
to form a rotationally supported disk/torus within the Bondi radius, i.e.,
\begin{equation}
R_{\rm Sch}\ll R_{\rm C}<R_{\rm B}.
\end{equation}
In this case, we find a global steady accretion solution which consists of three parts.
In the outer region ($r\sim R_{\rm B}$), a rotational equilibrium distribution is developed 
where the density distribution follows $\rho \propto (1+R_{\rm B}/r)^{3/2}$ and the gas motion is very subsonic.
The solution deviates only slightly from the rotational equilibrium flow with no radial motion 
\citep{Fishbone_Moncrief_1976}.
Interior to this ($r\la 2~R_{\rm C}$), a geometrically thick accretion torus is formed,
where thermal energy generated by viscosity is transported via strong convection motions.
Physical properties of the inner solution agree with those expected in a CDAF solution, e.g, 
$\rho \propto r^{-1/2}$ \citep{Quataert_2000,Narayan_2000}.
In the inner CDAF solution, the gas inflow rate decreases towards the center due to convection ($\dot{M}\propto r$),
and the net accretion rate (including both inflows and outflows) is strongly suppressed by several orders of magnitude from the Bondi accretion rate.
We also study the effect of the viscous strength on the results and find that 
the net accretion rate is approximated as $\dot{M}\sim (\alpha/0.01)^{0.6}(r/R_{\rm B})\dot{M}_{\rm B}$.
Finally, in a hot plasma at the bottom of this solution ($r\la 10^{-3}~R_{\rm B}$), 
thermal conduction would dominate the convective energy flux 
and the flow would resemble an optically thin accretion disk.
Since suppression of the accretion by convection ceases, the final BH feeding rate is determined 
as $\dot{M}\sim 10^{-3}-10^{-2}~\dot{M}_{\rm B}$.
This rate can explain why Sgr A$^*$ and the BHs in M31 and M87 accrete at 
such low accretion rates, without invoking any feedback mechanism.

The rest of this paper is organized as follows. 
In \S\ref{sec:method}, we describe the methodology of our numerical simulations. 
In \S\ref{sec:result}, we show our simulation results and explain their physical properties.
We compare the results to those in CDAF solutions and quantify the effects of energy and 
angular momentum transport by convection (\S\ref{sec:fid}), and study the effect of the choice of 
viscous parameters on the results (\S\ref{sec:vis}).
In \S\ref{sec:global}, we give an analytical argument extending the accretion solution further inwards and 
discuss a global solution of radiatively inefficient rotating accretion flows.
In \S\ref{sec:sum}, we summarize our main conclusions and discuss the importance for observations 
of very low-luminosity BHs.


\section{Methodology}
\label{sec:method}

\subsection{Basic quantities}

We introduce basic dimension-less physical quantities which characterize BH accretion systems.
If feedback by radiation and/or momentum from the central BH is negligible,
gas accretion begins from the Bondi radius, $R_{\rm B}$.
The standard accretion rate from the Bondi radius is given by
\begin{align}
\dot{M}_{\rm B}&=4\pi q(\gamma) \rho_\infty 
\frac{G^2M_\bullet ^2}{c_{\infty}^3},\nonumber\\
&\simeq 8.8\times 10^{-5}~\rho_{-22} M_6^2 c_7^{-3}~\msunyr,
\end{align}
where $q(\gamma)=1/4$ for $\gamma =5/3$, $\rho_{-22}=\rho_\infty/(10^{-22}~{\rm g~\cc})$,
 $M_6=M_\bullet/(10^6~\msun)$ and $c_7=c_\infty/(10^7~{\rm cm~s}^{-1})$ and
the accretion rate normalized by the Eddington rate is given by
\begin{align}
\dot{m}_{\rm B}\equiv \frac{\dot{M}_{\rm B}}{\dot{M}_{\rm Edd}}
=3.8\times 10^{-3}~\rho_{-22} M_6 c_7^{-3},
\end{align}
where $\dot{M}_{\rm Edd}(\equiv 10~L_{\rm Edd}/c^2) = 2.3\times 10^{-2}~M_6~\msunyr$ 
is the Eddington accretion rate with a $10\%$ radiative efficiency.

As shown in \S\ref{sec:intro}, there are three characteristic physical scales: 
the Bondi radius $R_{\rm B}$, the centrifugal radius $R_{\rm C}$ and 
the Schwarzschild radius $R_{\rm Sch}$.
From these three radii, we can define two ratios as 
\begin{equation}
\frac{R_{\rm Sch}}{R_{\rm B}}=\frac{2c_\infty^2}{c^2},
\end{equation}
and 
\begin{equation}
\frac{R_{\rm C}}{R_{\rm B}}=\frac{j^2c_\infty^2}{G^2 M_\bullet ^2}.
\label{eq:beta}
\end{equation}
Assuming a constant specific angular momentum of $j=\sqrt{\beta}R_{\rm B}c_\infty$,
the ratio is written as $R_{\rm C}/R_{\rm B}=\beta$.
Since we discuss cases where $R_{\rm Sch}\ll R_{\rm C}<R_{\rm B}$,
we assume $\beta<1$.

In axisymmetric two-dimensional hydrodynamical simulations, we need 
angular momentum transport via viscous processes in order to allow gas 
accretion onto a BH at the center.
We here model the effects of viscousity with the standard $\alpha$-prescription proposed by \cite{SS_1973}
instead of solving the time evolution of MHD equations.
According to MHD simulations, effective viscosity driven by the MRI provides angular momentum transport
and the strength is estimated as $\alpha \sim O(10^{-2})$
\citep{Balbus_Hawley_1991,Matsumoto_1995,Stone_1996,Balbus_Hawley_1998,Sano_2004}.

Throughout this paper, we focus on adiabatic gas without radiative cooling 
in order to keep numerical results scale-free.
The assumptions are justified for very low accretion-rate system, i.e., 
$\dot{M}_{\rm B}\ll \dot{M}_{\rm Edd}$ (see discussion \S\ref{sec:global}).
In this limit, we have only three important non-dimensional parameters of 
$R_{\rm Sch}/R_{\rm B}$, $R_{\rm C}/R_{\rm B}(=\beta)$ and $\alpha$ to define a BH accretion system.

\subsection{Basic equations}

We solve the axisymmetric two-dimensional hydrodynamical equations
using an open code {\tt PLUTO} \citep{Mignone_2007,Kuiper_2010,Kuiper_2011}.
The basic equations are following: the equation of continuity,
\begin{equation}
\frac{d\rho}{dt}+\rho \nabla \cdot \mbox{\boldmath $v$}=0,
\end{equation}
and the equation of motion,
\begin{equation}
\rho \frac{d\mbox{\boldmath $v$}}{dt}=-\nabla p -\rho \nabla \Phi 
+ \nabla \cdot \mbox{\boldmath $\sigma$},
\end{equation}
where $\rho$ is the density, {\boldmath $v$} is the velocity, and 
$p$ is the gas pressure,
the gravitational potential is set to $\Phi =-GM_\bullet /r$, 
and {\boldmath $\sigma$} is the stress tensor due to viscosity.
The time derivative is the Lagrangian derivative, given by 
$d/dt\equiv \partial/\partial t$ + {\boldmath $v$}$ \cdot \nabla$.

We solve the energy equation of
\begin{equation}
\rho \frac{de}{dt}=-p\nabla \cdot \mbox{\boldmath $v$} 
+ (\mbox{\boldmath $\sigma$} \cdot \nabla) \mbox{\boldmath $v$},
\end{equation}
where $e$ is the internal energy per mass.
The equation of state of the ideal gas is assumed as $p=(\gamma-1)\rho e$, 
where $\gamma = 1.6$\footnote{
For a spherically symmetric flow, the sonic point is estimated as 
$R_{\rm sonic}=[(5-3\gamma)/4]R_{\rm B}$.
Since for $\gamma=5/3$, the sonic point is located at the origin,
the inner boundary conditions would affect gas dynamics in
the computational domain.
Thus, we set a slightly smaller value of $\gamma=1.6$, for which 
the sonic point can be resolved ($R_{\rm sonic}\simeq 0.05~R_{\rm B}$).}.
The two terms on the right-hand-side present compressional heating 
(or expansion cooling) and viscous heating.

The viscous stress tensor is given by
\begin{equation}
\sigma_{ij}=\rho \nu \left[ 
\left(\frac{\partial v_j}{\partial x_i} + \frac{\partial v_i}{\partial x_j}\right)
-\frac{2}{3} (\nabla \cdot \mbox{\boldmath $v$} )\delta_{ij}
\right],
\end{equation}
where $\nu$ is the shear viscosity.
Note that the bulk viscosity is neglected here.
The shear viscosity is calculated with the $\alpha$-prescription \citep{SS_1973},
\begin{equation}
\nu = \alpha \frac{c_{\rm s}^2}{\Omega_{\rm K}},
\label{eq:alpha_visc}
\end{equation}
where 
$\Omega_{\rm K}=(GM_\bullet/r^3)^{1/2}$.
We calculate the viscous parameter by mimicking some properties of the MRI as
\begin{equation}
\alpha = \alpha_0 \exp{\left[ 
-\left(\frac{\rho_{\rm cr}}{\rho}\right)^2\right]},
\label{eq:alpha}
\end{equation}
where $\rho_{\rm cr}$ is a threshold of the density, above which viscosity turns on.
We adopt a maximum value of the density at an initial condition as the threshold value (see \S\ref{sec:ic}).
Under this model, the viscous process is active almost only within an accretion disk ($r\la R_{\rm C}$),
where the rotational velocity has a significant fraction of the Keplerarian value.
On the other hand, since the viscosity becomes zero outside the disk, angular momentum transported 
from the disk tends to be accumulated outside it 
as shown by the ``bump" in the angular momentum distribution in Fig.~\ref{fig:app_2}, 
where no viscous processes operate.
Exterior to the peak in the angular momentum distribution where
the specific angular momentum has a negative 
gradient outward, i.e., $\partial j/\partial r <0$, so-called Rayleigh's criterion.
In reality, such rotating flows are unstable and become turbulent \citep[e.g.,][]{Chandrasekhar_1961}, 
leading to angular momentum 
transport in three-dimensional simulations.
However, because of limitations of our two-dimensional simulations, 
this could not occur and angular momentum flowing out from the central regions would accumulate outside 
of the region where $\rho/\rho_{\rm cr}\sim 1$.
Thus, we modify Eq. (\ref{eq:alpha}) by adding the second term as 
\begin{equation}
\alpha = \alpha_0 \left\{ \exp{\left[ 
-\left(\frac{\rho_{\rm cr}}{\rho}\right)^2\right]} + {\rm max}\left(0, -\frac{\partial \ln j}{\partial \ln r}\right)\right\},
\label{eq:alpha_j}
\end{equation}
In a physical sense, the second term is necessary so that a steady state of the accretion flow exists.
Note that the treatment of the rotational instability does not affect our results (see Appendix \ref{sec:app}).

\subsection{Boundary and initial conditions}
\label{sec:ic}

We set a computational domain of $r_{\rm min} \leq r \leq r_{\rm max}$ 
and $\epsilon \leq \theta \leq \pi -\epsilon$, where $\epsilon$ is set to 
$0.01$ radian to avoid numerical singularity at poles.
We set logarithmically-spaced grids in the radial direction
and uniformly-spaced grids in the polar direction. 
The number of the grid points is set to $(N_r, N_\theta) = (512, 256)$.
We have checked convergence for our simulation results, changing the number of the grids
(see Appendix \ref{sec:app}).
As our fiducial case, we set $r_{\rm min}=7.7\times 10^{-3}~R_{\rm B}$ and $r_{\rm max}=54~R_{\rm B}$.
Moreover, we calculate two cases with $r_{\rm min}=3.0\times 10^{-3}~R_{\rm B}$ and $1.3\times 10^{-2}~R_{\rm B}$
(see \S\ref{sec:ang} and \S\ref{sec:vis}).
Note that the relative size of each cell is given by $\Delta r/r=(r_{\rm max}/r_{\rm min})^{1/N_r}$.

We adopt the outflow boundary condition at the innermost/outermost grid \citep[e.g.,][]{Stone_Norman_1992},
where zero gradients cross the boundary are imposed on physical quantities
in order to avoid spurious reflection of wave energy at the boundary.
We also impose $v_r\leq 0$ at the inner boundary 
(i.e., inflowing gas from ghost cells is prohibited).
At the poles ($\theta=\epsilon$, $\pi-\epsilon$), the reflective condition is imposed on
the circumferential component of the velocity $v_\theta$.

As initial conditions, we adopt a rotational equilibrium distribution ($v_r=v_\theta=0$) 
with a constant specific angular momentum of $j_\infty$,
\begin{equation}
\frac{\rho}{\rho_\infty}=\left[1+(\gamma-1)\frac{GM_\bullet}{c_\infty^2 r} 
- \frac{(\gamma-1)}{2}\frac{j_\infty^2}{c_{\rm \infty}^2\varpi^2} \right]
^{\frac{1}{\gamma-1}}
\label{eq:torus}
\end{equation}
(\citealt{Papaloizou_Pringle_1984}; see also \citealt{Fishbone_Moncrief_1976}),
where $\varpi = r\sin \theta $ is the cylindrical radius.
Hereafter, we refer to this as the ``Fishbone-Moncrief" solution.
The first and second terms on the right-hand side present density enhancement via gravity of the central BH
inside the Bondi radius.
The third term expresses the centrifugal force for a given $j_\infty$,
which leads to a maximum value of the density at $r=R_{\rm C}$ and $\theta =0$
\begin{equation}
\rho_{\rm cr}=\rho_\infty \left(1+\frac{\gamma-1}{2\beta}\right)^{1/(\gamma-1)},
\end{equation}
where $\beta=R_{\rm C}/R_{\rm B}$ as defined below Eq. (\ref{eq:beta}).
Without viscosity, the density never exceeds this value because of the centrifugal barrier.
In other words, a high-density region with $\rho>\rho_{\rm cr}$ must be formed by 
angular momentum transport due to viscosity.

In order to study the dependence of our numerical results on the choice of the initial conditions,
we also conduct a simulation which starts from the Bondi accretion solution
with a constant angular momentum $j_\infty$.
Note that the numerical result with the Bondi profiles as initial conditions
approaches that adopting a rotational equilibrium distribution given by Eq. (\ref{eq:torus})
(see Appendix \ref{sec:app}).

\section{Results}
\label{sec:result}

In this section, we show results of our two-dimensional simulations for accretion flows.
We set all physical quantities as $M_\bullet=10^6~\msun$, $\rho_\infty = 10^{-22}~{\rm g~\cc}$, 
and $c_\infty=100~\kms$ ($\dot{m}_{\rm B}\simeq 3.8\times 10^{-3}$).
As already mentioned, in adiabatic cases, our results do not depend on $\dot{m}_{\rm B}$ 
but on the two dimensionless parameters of $\beta =R_{\rm C}/R_{\rm B}$ and $\alpha$.

In \S\ref{sec:fid}, we first describe overall properties of accretion flows with 
different initial angular momentum $\beta$ for our fiducial case with $\alpha=0.01$.
In \S\ref{sec:vis}, the dependence of our results on the choice of viscous parameter is discussed.

\subsection{Fiducial case with $\alpha=0.01$}
\label{sec:fid}

\begin{figure}
\begin{center}
\includegraphics[width=82mm]{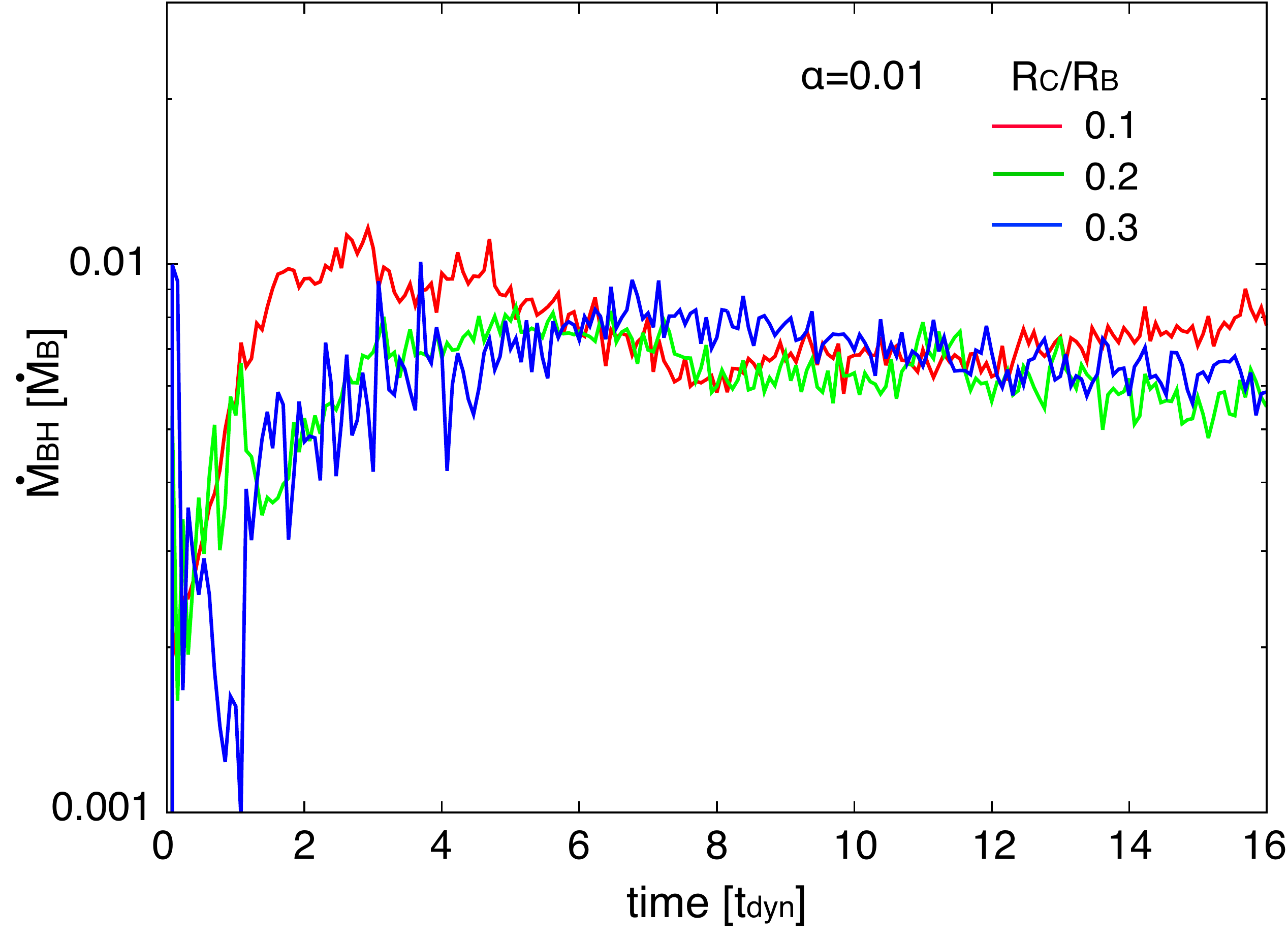}
\caption{Time evolution of the net accretion rate (in units of the Bondi rate) onto a BH, 
i.e., a sink cell for different values of angular momentum: 
$R_{\rm C}/R_{\rm B}=0.1$ (red), $0.2$ (green) and $0.3$ (blue).
The viscous parameter is set to $\alpha =0.01$.
The time-averaged accretion rates ($t>6~t_{\rm dyn}$) are 
$\dot{M}\simeq 7\times 10^{-3}~\dot{M}_{\rm B}$.
}
\label{fig:t_Mdot_j}
\end{center}
\end{figure}

Fig.~\ref{fig:t_Mdot_j} shows the time evolution of gas accretion rates onto a BH
(i.e., a sink cell at $r=r_{\rm min}$) for different initial values of angular momentum:
$R_{\rm C}/R_{\rm B}=0.1$ (red), $0.2$ (green) and $0.3$ (blue).
The viscous parameter is set to $\alpha =0.01$.
We normalize the simulation time by the dynamical timescale at the Bondi radius,
$t_{\rm dyn}(\equiv R_{\rm B}/c_\infty) \simeq 1.3\times 10^{11}~M_6 c_7^{-3}~{\rm s}$.
For all the cases, the accretion rates increase with fluctuations at $t\la 4~t_{\rm dyn}$
and saturate at almost constant values by $t\simeq 5~t_{\rm dyn}$.
The behavior of the accretion rates in all the cases is similar.
In fact, the time-averaged values over $6 \leq t/t_{\rm dyn} \leq 16$ are 
$\dot{M}/\dot{M}_{\rm B}\simeq 7\times 10^{-3}$, which does not depend on 
the initial angular momentum.
The results show that gas rotation reduces the inflow rates by two orders of magnitude from the Bondi rate.
Note that the suppression factor of the accretion rate is close to the viscous parameter we assume ($\alpha =0.01$).
Of course in the limit of $\alpha=0$, there is no net accretion.

\begin{figure}
\begin{center}
\includegraphics[width=82mm]{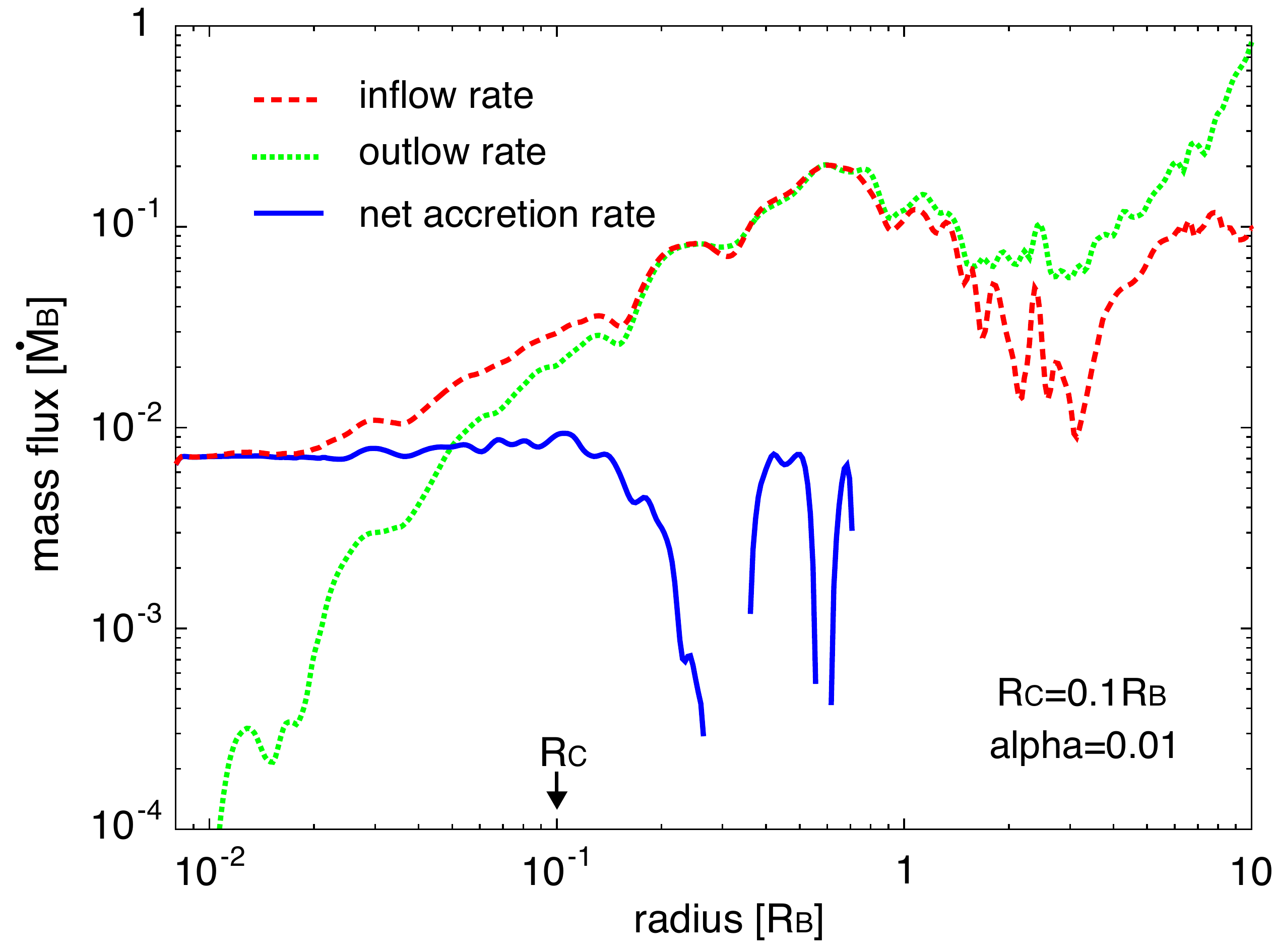}
\caption{Radial structure of the angle-integrated mass inflow and outflow rates
for the case with $R_{\rm C}/R_{\rm B}=0.1$ and $\alpha=0.01$ at $t\simeq 13~t_{\rm dyn}$.
Each curve presents the inflow rate ($-\dot{M}_{\rm in}$ in Eq. \ref{eq:Min}; red), 
the outflow rate ($\dot{M}_{\rm out}$ in Eq. \ref{eq:Mout}; green),
and the net accretion rate ($-\dot{M}\equiv -\dot{M}_{\rm in}-\dot{M}_{\rm out}$; blue), respectively.
Inside the Bondi radius ($r\la R_{\rm B}$), the inflow and outflow rate decrease towards the center 
following $\propto r$ and are tightly balanced.
Within the centrifugal radius, where a geometrically-thick accretion disk forms, 
the net accretion rate becomes nearly constant.
}
\label{fig:r_Mdot}
\end{center}
\end{figure}

Fig.~\ref{fig:r_Mdot} shows the radial structure of the angle-integrated 
mass inflow (red) and outflow (green) rates for the case with $R_{\rm C}/R_{\rm B}=0.1$ and
$\alpha=0.01$ at $t\simeq 13~t_{\rm dyn}$.
Those rates are defined by
\begin{align}
\dot{M}_{\rm in}(r)&=2\pi r^2\int \rho \cdot {\rm min}(v_r,0)\sin \theta d\theta,
\label{eq:Min}\\
\dot{M}_{\rm out}(r)&=2\pi r^2\int \rho \cdot {\rm max}(v_r,0)\sin \theta d\theta,
\label{eq:Mout}
\end{align}
where $\dot{M}_{\rm in}< 0$ and $\dot{M}_{\rm out}> 0$ \citep[e.g.,][]{Stone_1999}.
Note that $\dot{M}_{\rm out}$ in Eq. (\ref{eq:Mout}) does not necessarily mean 
the outflow rates of gas escaping to the infinity, but takes account into gas flows with $v_r>0$.
We also show the net accretion rate defined by $-\dot{M}\equiv -\dot{M}_{\rm in}-\dot{M}_{\rm out}$ (blue).
Inside the Bondi radius ($r\la R_{\rm B}$), both inflows and outflows exist and those rates,
which decrease towards the center following $\propto r$, are tightly balanced.
Within the centrifugal radius ($r\la R_{\rm C}=0.1~R_{\rm B}$),
the net accretion rate becomes a constant value ($\simeq 7\times 10^{-3}~\dot{M}_{\rm B}$).
We note that the outflow rate dominates the inflow rate outside the Bondi radius.
In Appendix \ref{sec:app}, we study the time evolution of the outflowing gas and 
make sure that this component does not affect stationarity of the accretion flow (see in Fig.~\ref{fig:outflow}).

\begin{figure*}
\begin{center}
\includegraphics[width=140mm]{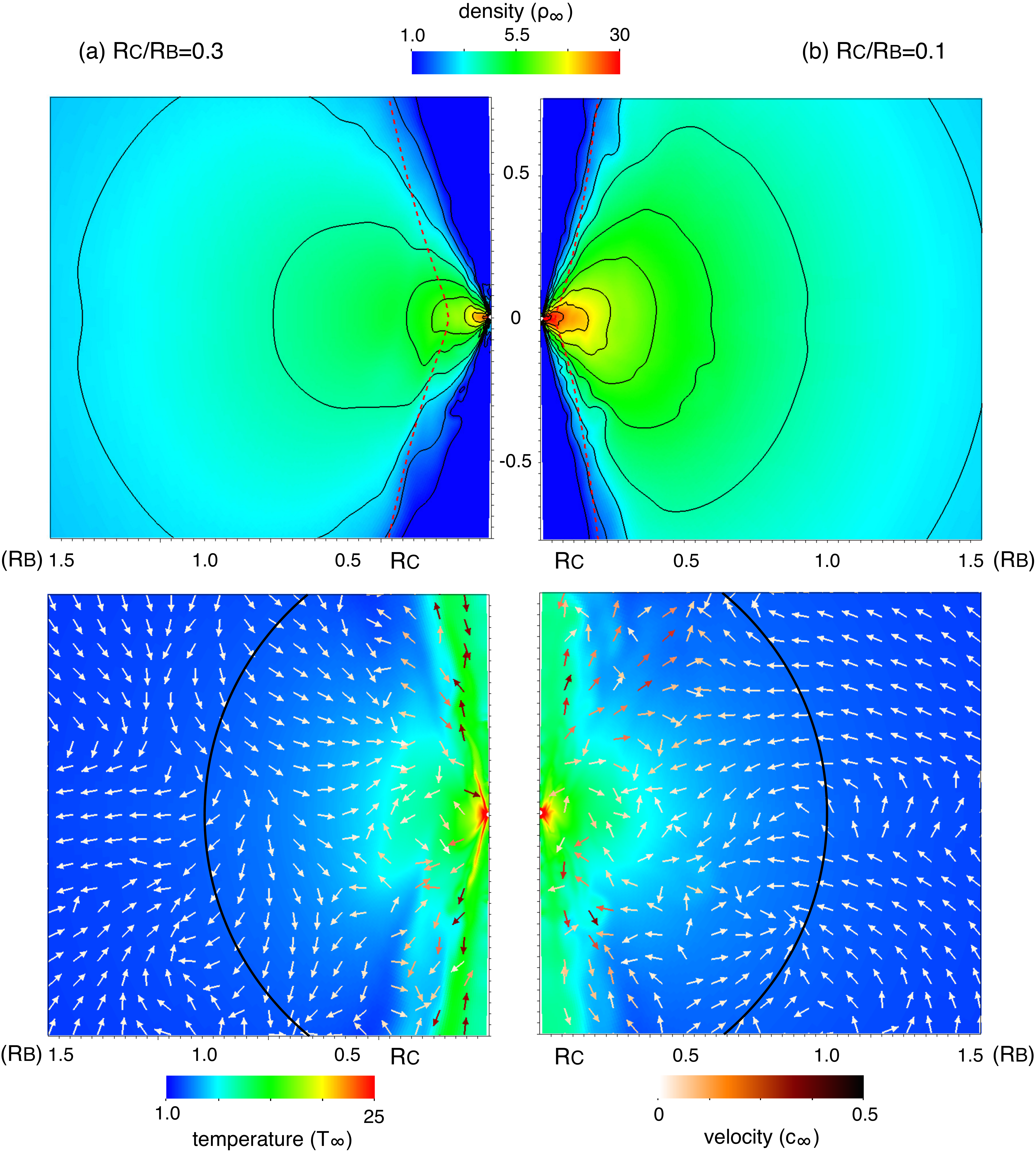}
\caption{
Distribution of the gas density (top) and temperature (bottom)
for different values of angular momentum with $R_{\rm C}/R_{\rm B}=0.3$ (left) and $0.1$ (right).
Isodensity contours and velocity vectors are shown in the top and bottom panel, respectively.
The elapsed time is set to $t=13~t_{\rm dyn}$ as shown in Fig.~\ref{fig:r_Mdot}.
In the top panel, the zero-density boundary for the initial distribution is shown 
by red dashed curve, inside which the density distribution deviates from the equilibrium solution.
The thick black curve in the bottom panel shows the location of the Bondi radius.
}
\label{fig:rho_temp_ad}
\end{center}
\end{figure*}

In Fig.~\ref{fig:rho_temp_ad}, we present the distribution of the gas density (top) and temperature (bottom)
for two cases with $R_{\rm C}/R_{\rm B}=0.3$ (left) and $0.1$ (right).
We also plot isodensity contours and the velocity vectors in the top and bottom panel, respectively.
The elapsed time is set to $t=13~t_{\rm dyn}$ as shown in Fig.~\ref{fig:r_Mdot}.
For two cases, the density distribution around the Bondi radius is approximated
by a rotational equilibrium distribution with a constant specific angular momentum,
i.e., the Fishbone-Moncrief solution (see Eq. \ref{eq:torus} and Fig.~\ref{fig:r_rho} below).
Red dashed curve presents the boundary where $\rho=0$ for the initial distribution.
At the vicinity of the centrifugal radius ($r\la 2~R_{\rm C}$) and inside the zero-density boundary 
for the initial conditions (red dashed), the density distribution deviates 
from the equilibrium solution since the angular momentum begins to be transported outward
and the profile approaches $j\propto r^{1/2}$ (see Figs.~\ref{fig:r_vel} and \ref{fig:app_2} below).
The gas temperature increases toward the center by compressional heating due to the gravity of the BH
and energy dissipation due to viscosity.
Gas motions in non-azimuthal directions are subsonic and form circulations.
We note that some fraction of the gas is outflowing from inside the Bondi radius (thick black curve).
In the polar directions, weakly collimated hot outflows are launched 
(see also green curve in Fig.~\ref{fig:r_Mdot} at $r\ga 2~R_{\rm B}$).
The outflow speed increases with the angular momentum of the system.

\begin{figure*}
\begin{center}
\includegraphics[width=140mm]{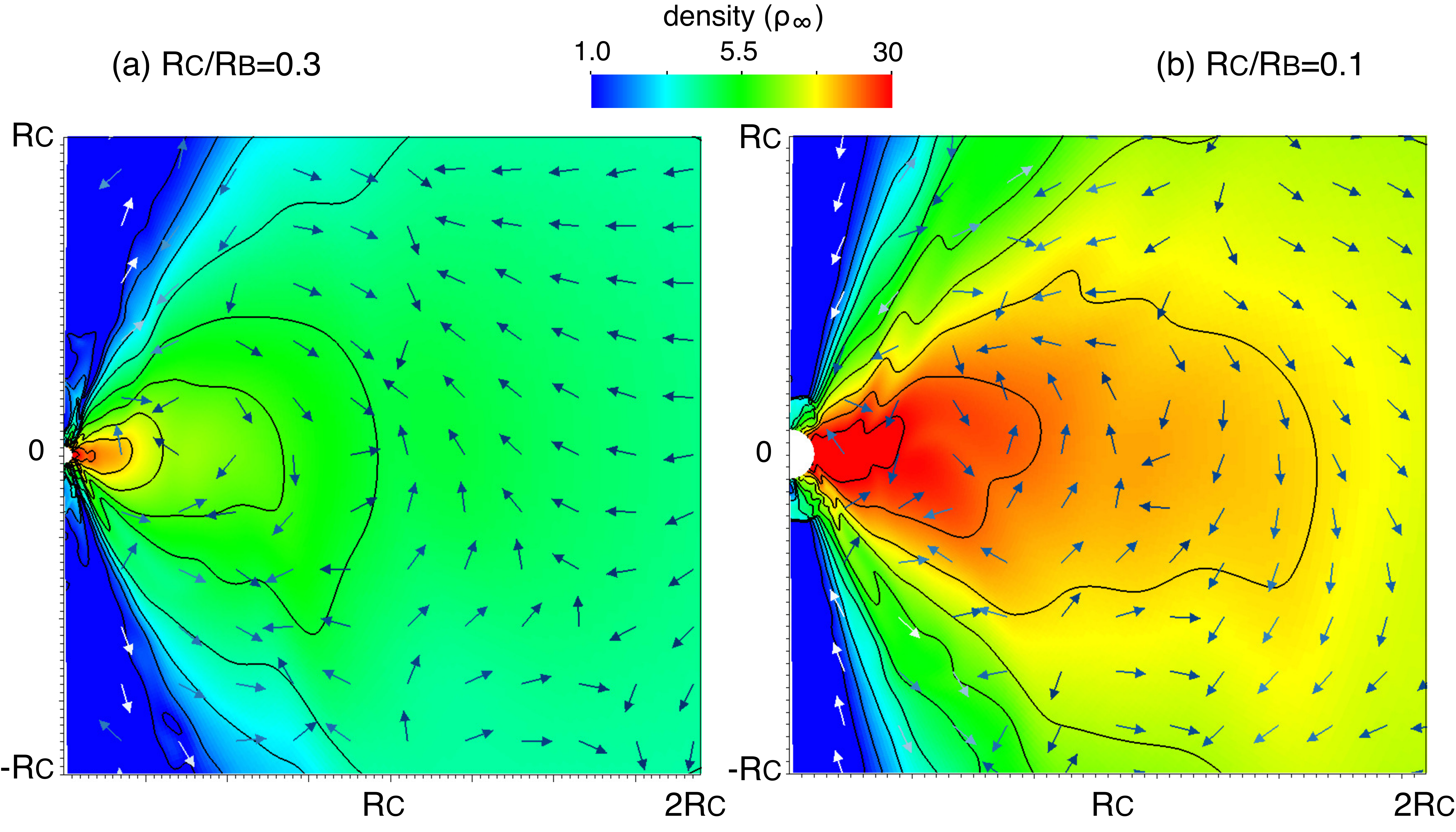}
\caption{
Distribution of the gas density in the inner region
for different values of angular momentum with $R_{\rm C}/R_{\rm B}=0.3$ (left) and $0.1$ (right).
Isodensity contours and velocity vectors are shown, and the elapsed time is set to $t=13~t_{\rm dyn}$ 
as shown in Fig.~\ref{fig:r_Mdot}.
}
\label{fig:rho_temp_ad_small}
\end{center}
\end{figure*}

Fig.~\ref{fig:rho_temp_ad_small} shows the density distribution at smaller scale
for $R_{\rm C}/R_{\rm B}=0.3$ (left) and $0.1$ (right), respectively.
For the two cases, a dense torus structure is formed around the central BH, 
where the gas motions are very subsonic even inside the Bondi radius and 
highly convective (see \S\ref{sec:ss}).
From the torus surface with $\rho \simeq 2~\rho_\infty$, outflows are launched towards the poles.
Except in the polar directions, the gas is both inflowing and outflowing, and the two rates are almost balanced.
We emphasize that mirror-symmetry of the accretion flow cross the equatorial plane ($\theta =\pi/2$) is broken.
Although a previous study by \cite{LOS_2013} imposed equatorial mirror-symmetry and found that  
a large fraction of the gas is outflowing through the equator coherently, 
the equatorial outflow is an artefact of the imposed symmetry (see also \citealt{Roberts_2017}).

Angular momentum affects the gas density near the central hole.
Namely, the absolute value is higher with lower angular momentum.
This is because concentration of the density is suppressed by convective motions
inside the centrifugal radius.
As we will discuss in \S\ref{sec:ss}, the density profile becomes flatter ($\rho \propto r^{-1/2}$) 
inside the convective envelope ($r\la 2~R_{\rm C}$) rather than $\rho \propto r^{-3/2}$.

\subsubsection{Comparison to self-similar solutions}
\label{sec:ss}

In order to understand properties of the accretion flows,
we consider time-averaged profiles of physical quantities 
and compare to those of self-similar solutions for radiatively inefficient accretion flows.
In the following, we show time-averaged values over $6 \leq t/t_{\rm dyn} \leq 16$.

\begin{figure}
\begin{center}
\includegraphics[width=82mm]{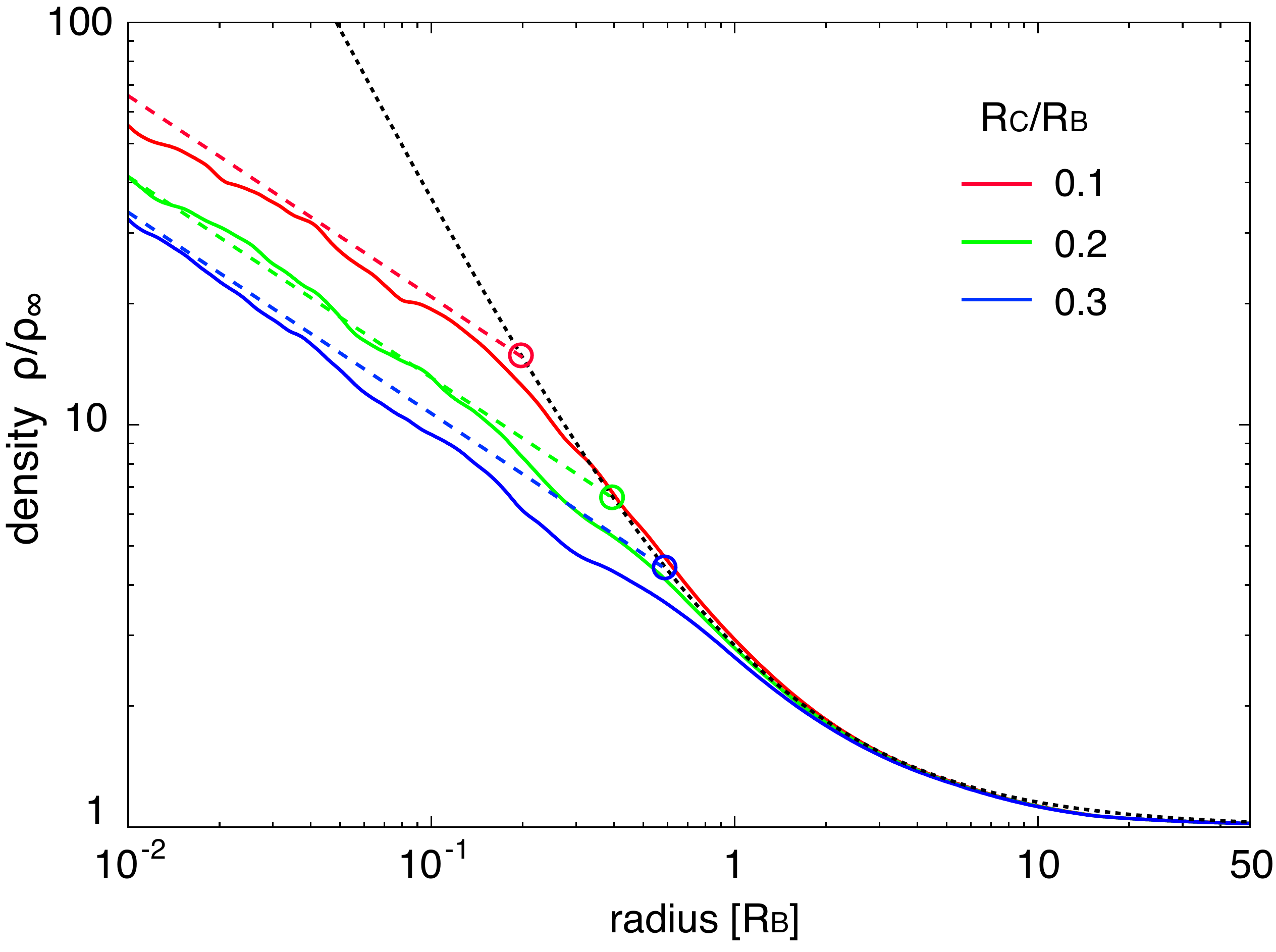}
\caption{Radial profiles of the gas density along the equator ($\theta =\pi/2$) for 
the cases with $R_{\rm C}/R_{\rm B}=0.1$ (red) $0.2$ (green) and $0.3$ (blue).
The density profiles are time-averaged over $6\leq t/t_{\rm dyn} \leq 16$,
and consist of two components.
In the outer region, they are approximated by an equilibrium solution of rotating gas, 
``Fishbone-Moncrief solution", following $\rho/\rho_\infty=(1+R_{\rm B}/r)^{3/2}$ (dotted curve).
In the inner region, the density profiles approach $\rho \propto r^{-1/2}$ 
(dashed lines; Eq. \ref{eq:rho_innner}), whose slope is the same as in CDAF solutions.
The transition occurs at $r\simeq 2~R_{\rm C}$ for each case (open circles).
}
\label{fig:r_rho}
\end{center}
\end{figure}

\begin{figure}
\begin{center}
\includegraphics[width=75mm]{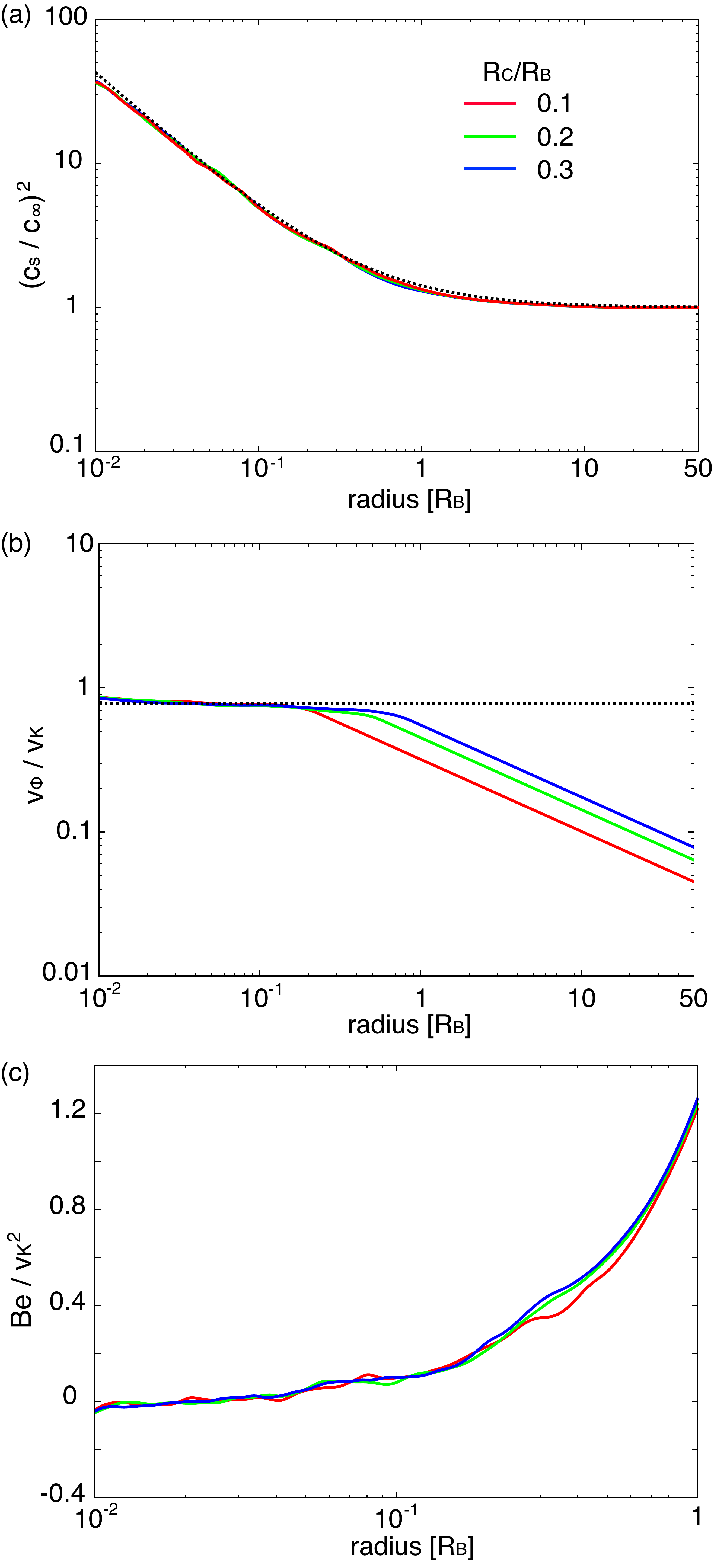}
\caption{Radial profile of the time-averaged (a) sound speed, (b) rotational velocity and (c) Bernoulli number
along the equator ($\theta =\pi/2$) for three different cases with $R_{\rm C}/R_{\rm B}=0.1$ (red), 
$0.2$ (green) and $0.3$ (blue).
The dotted curves in the top two panels (Eqs. \ref{eq:an1} and \ref{eq:an3}) are those of the CDAF solutions, respectively.
}
\label{fig:r_vel}
\end{center}
\end{figure}

Fig.~\ref{fig:r_rho} shows radial profiles of the gas density along the equator ($\theta=\pi/2$)
for three cases with $R_{\rm C}/R_{\rm B}=0.1$ (red), $0.2$ (green) and $0.3$ (blue).
The profiles consist of two components. 
In the outer region ($r> 2~R_{\rm C}$), the density profiles are similar to the equilibrium solution
of $\rho/\rho_\infty=(1+R_{\rm B}/r)^{3/2}$ (black dotted).
On the other hand, the density follows $\rho \propto r^{-1/2}$ in the inner region ($r< 2~R_{\rm C}$)
and is well approximated by
\begin{equation}
\rho \simeq \rho_\infty \frac{(1+2\beta)^{3/2}}{2\beta}\left(\frac{r}{R_{\rm B}}\right)^{-1/2},
\label{eq:rho_innner}
\end{equation}
(see dashed lines in Fig.~\ref{fig:r_rho}).
The slope of the density profile in this region agrees with that for a CDAF solution 
($\rho \propto r^{-1/2}$; \citealt{Quataert_2000, Narayan_2000})
rather than that for an ADAF solution ($\rho \propto r^{-3/2}$; \citealt{NY_1994,NY_1995}).

Fig.~\ref{fig:r_vel} shows radial profiles along the equator of (a) the sound speed, 
(b) the rotational velocity and (c) the Bernoulli number, which is defined by
\begin{equation}
Be \equiv \frac{v^2}{2}+\frac{c_{\rm s}^2}{\gamma-1}-\frac{GM_\bullet}{r}.
\end{equation}
Those profiles inside the centrifugal radius ($r<2~R_{\rm C}$) are summarized as 
$c_{\rm s}^2\propto r^{-1}$, $v_\phi/v_{\rm K}\simeq 1$, and $Be/v_{\rm K}^2\simeq 0$,
where $v_{\rm K}=\sqrt{GM_\bullet/r}$ is the Keplerian velocity.
Since gas accretion begins from the Bondi radius, the Bernoulli number is close to zero 
unless energy dissipation via viscosity is significant.
Within the centrifugal radius ($r\la R_{\rm C}$), the rotational velocity dominates and is approximated as $\sim v_{\rm K}$.
Therefore, the sound speed follows as $c_{\rm s}^2\propto r^{-1}$ as shown in panel (a).
We note that the profiles of the sound speed do not depend on the angular momentum of the flow.
In fact, the profiles can be fit with 
\begin{equation}
c_{\rm s}^2=c_{\infty}^2 \left[ 1+f(\gamma)\frac{R_{\rm B}}{r} \right],
\label{eq:an1}
\end{equation}
where we adopt a functional form of $f(\gamma)$ as
\begin{equation}
f(\gamma)\equiv \frac{\gamma-1}{\gamma+1-(\gamma-1)/2}
\label{eq:an2}
\end{equation}
Moreover, the rotational velocity inside the centrifugal radius is approximated as
\begin{equation}
\frac{v_\phi}{v_{\rm K}} = g(\gamma)  \equiv  \sqrt{\frac{2-(\gamma-1)}{\gamma+1-(\gamma-1)/2}}.
\label{eq:an3}
\end{equation}
For $\gamma=1.6$, $f(\gamma)=0.26$ and $g(\gamma)=0.78$.
These functions of $f$ and $g$ have been calculated by \cite{Quataert_2000}
for marginally stable rotating accretion flows against convection motions, i.e., CDAF solutions.

\begin{figure}
\begin{center}
\includegraphics[width=82mm]{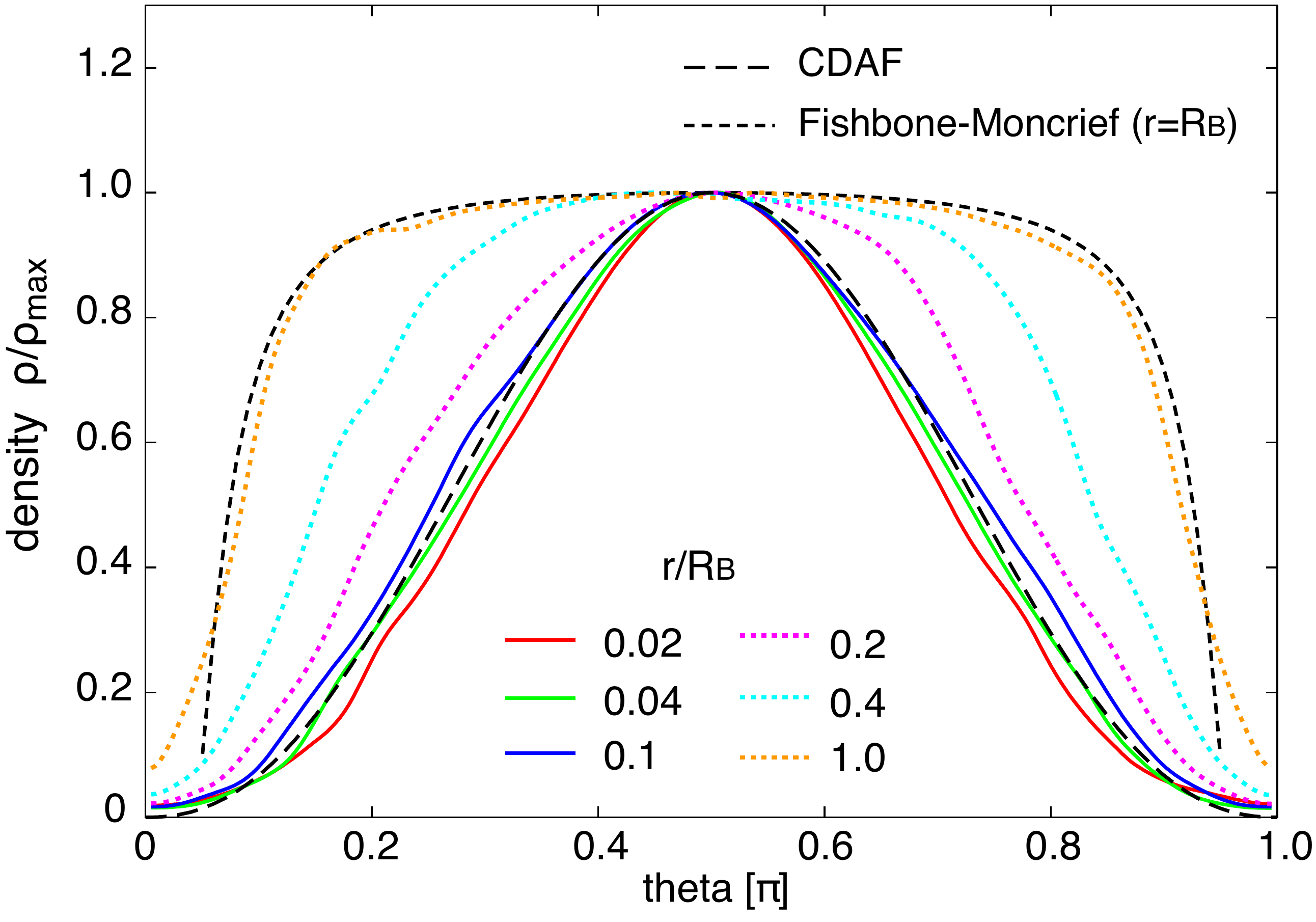}
\caption{Angular profiles of the time-averaged density at radial positions of
$0.02\leq r/R_{\rm B}\leq 1.0$.
The results in the inner region ($r<2~R_{\rm C}$) and outer region ($r\geq 2R_{\rm C}$)
are shown by solid and dotted curves, respectively.
Analytical angular profiles for CDAF solutions $\rho(\theta)\propto (\sin \theta)^{[2/(\gamma-1)-1]}$ (long dashed)
and for the Fishbone-Moncrief solution at $r=R_{\rm B}$ (short dashed) are shown for comparison to the numerical results.
The profiles are normalized by the maximum value for each case.
}
\label{fig:theta_rho}
\end{center}
\end{figure}

\begin{figure}
\begin{center}
\includegraphics[width=82mm]{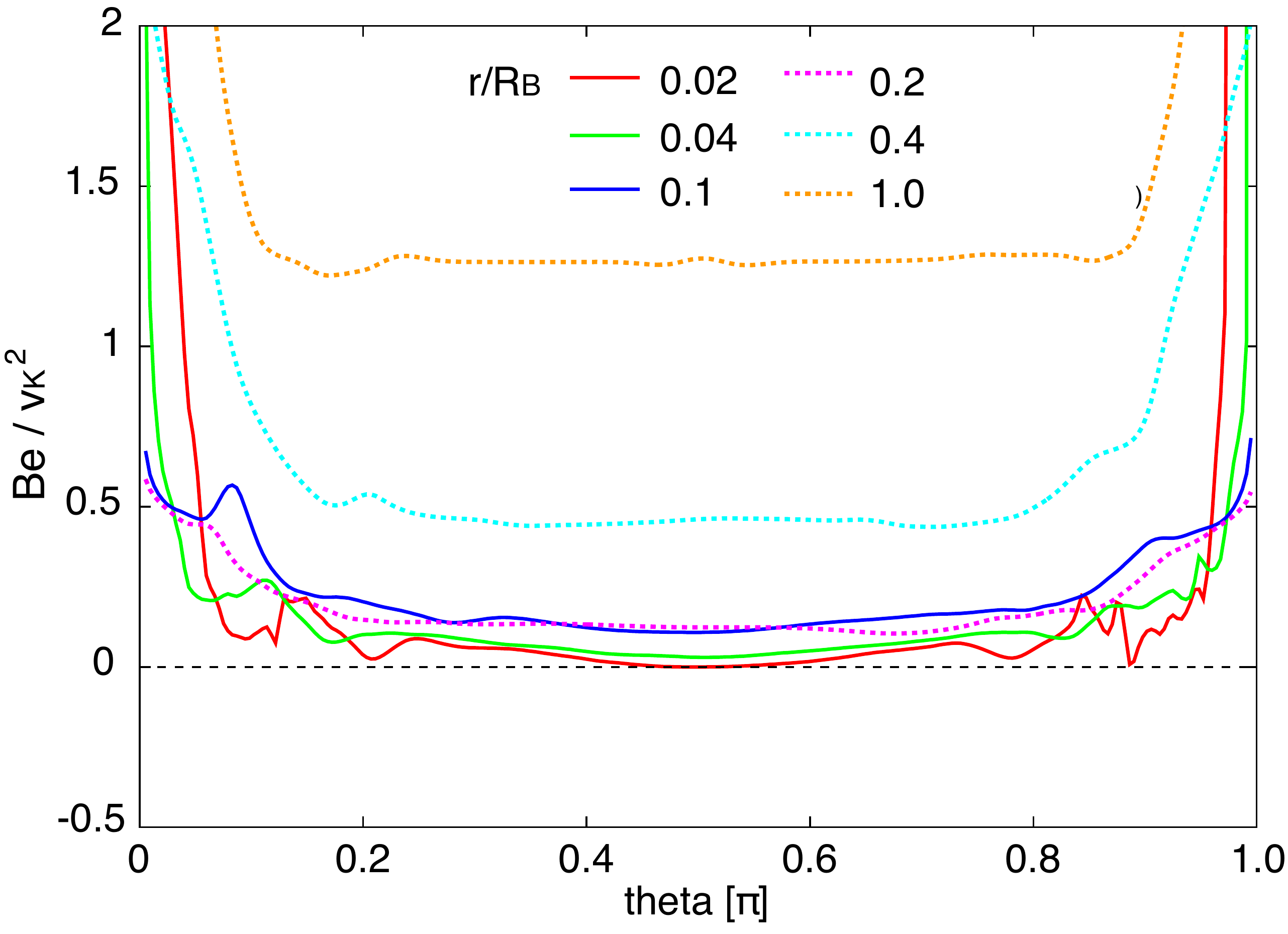}
\caption{Angular profiles of the time-averaged Bernoulli number at radial positions of
$0.02\leq r/R_{\rm B}\leq 1.0$.
In the inner region ($r<2~R_{\rm C}$), the value of $Be$ is smaller than $\sim 0.2~v_{\rm Kep}^2$.
Note that the value increases rapidly towards the poles, where weak outflows are launched.
}
\label{fig:theta_Be}
\end{center}
\end{figure}

Fig.~\ref{fig:theta_rho} shows profiles of the density as a function of the polar angle $\theta$,
at radial positions of $0.02\leq r/R_{\rm B}\leq 1.0$.
In the outer region ($r\ga 2R_{\rm C}$), the density profile follows an equilibrium distribution
and explained by the Fishbone-Moncrief solution (short-dashed).
In the inner region ($r<2~R_{\rm C}=0.2~R_{\rm B}$), the density concentration to the mid-plane is higher. 
The angular profiles in the inner region can be explained by long-dashed curve,
$\rho(\theta)\propto (\sin \theta)^{2[1/(\gamma-1)-1/2]} [=(\sin \theta)^{2.33}~{\rm for ~\gamma=1.6}]$
\citep{Quataert_2000}.
This result also indicates that the profiles of the accretion flow within the centrifugal radius 
are self-similar, and those properties approach those in CDAFs.
Fig.~\ref{fig:theta_Be} also presents angular profiles of the Bernoulli number.
In the inner region ($r<2~R_{\rm C}$), the value of $Be$ is smaller than $\la 0.2~v_{\rm Kep}^2$
except in the vicinity of the poles.
We note that the value increases rapidly towards the poles, where weak outflows are launched.

\subsubsection{Angular momentum and energy transport via convection motions}
\label{sec:ang}

\begin{figure}
\begin{center}
\includegraphics[width=82mm]{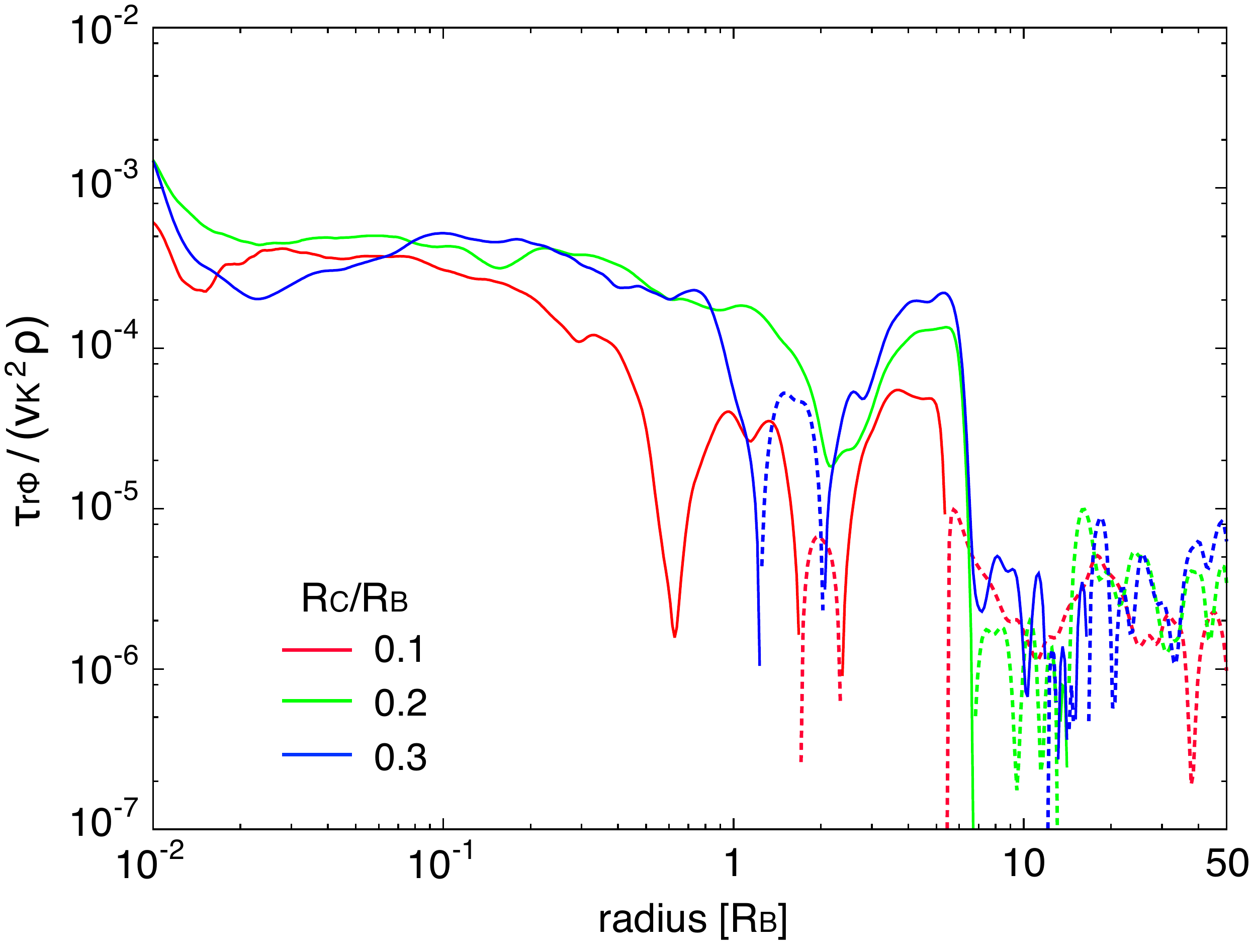}
\caption{Radial profiles of the $r-\phi$ component of the Reynolds stress $\tau_{r\phi}$ for different three cases with
$R_{\rm C}/R_{\rm B}=0.1$ (red), $0.2$ (green) and $0.3$ (blue).
Solid and dashed curves show positive and negative values, respectively.
The values are normalized by $\rho v_{\rm K}^2$.
Since $\tau_{r\phi}\propto Ar^{-3/2}$ is positive, convection motions transports 
angular momentum inward within the centrifugal radius.
}
\label{fig:Rey}
\end{center}
\end{figure}

Angular momentum transport by convection has been discussed with analytical methods 
and numerical simulations by previous works 
\citep[e.g.,][]{Quataert_2000, Narayan_2000,IA_2000,IAN_2000,INA_2003}.
To analyze the effect, we calculate the $r$-$\phi$ component of Reynolds stress,
$\tau_{r \phi}\equiv -\rho \langle v'_r v'_\phi \rangle$, where $v'_i\equiv v_i-\langle v_i \rangle$ and 
$\langle \cdot \rangle$ means the time-averaged value over $6\leq t/t_{\rm dyn}\leq 16$.
In practice, the mass-weighted Reynolds stress integrated over the polar angle is calculated as 
\begin{equation}
\tau_{r\phi}(r)
= -\int ^\pi_0 \langle \rho v'_r v'_\phi \rangle \sin \theta d \theta.
\end{equation}
In Fig.~\ref{fig:Rey}, we show the radial profile of the Reynolds stress normalized by $\rho v_{\rm K}^2$
for three cases of $R_{\rm C}/R_{\rm B}=0.1$ (red), $0.2$ (green) and $0.3$ (blue).
The Reynolds stress does not depend on the initial angular momentum,
and the value is approximated as $\tau_{r\phi}\simeq Ar^{-3/2}$, where $A$ is positive.
Note that the $r$-$\phi$ component of the viscous stress has a negative value (i.e., $\sigma_{r\phi}r^{3/2}\simeq {\rm const.}<0$).
Therefore, the positive sign of $A$ means that convective motions transport angular momentum inward,
while the standard $\alpha$-viscosity transports it outward. 
\cite{IAN_2000} have conducted three-dimensional simulations relaxing the assumption of axisymmetry 
and found that the results are essentially similar to those obtained in two-dimensional simulations.
In fact, the convective eddies are nearly axisymmetric and transport angular momentum inward.
This allows us to justify that our two-dimensional simulations can capture the important physics.

\begin{figure}
\begin{center}
\includegraphics[width=82mm]{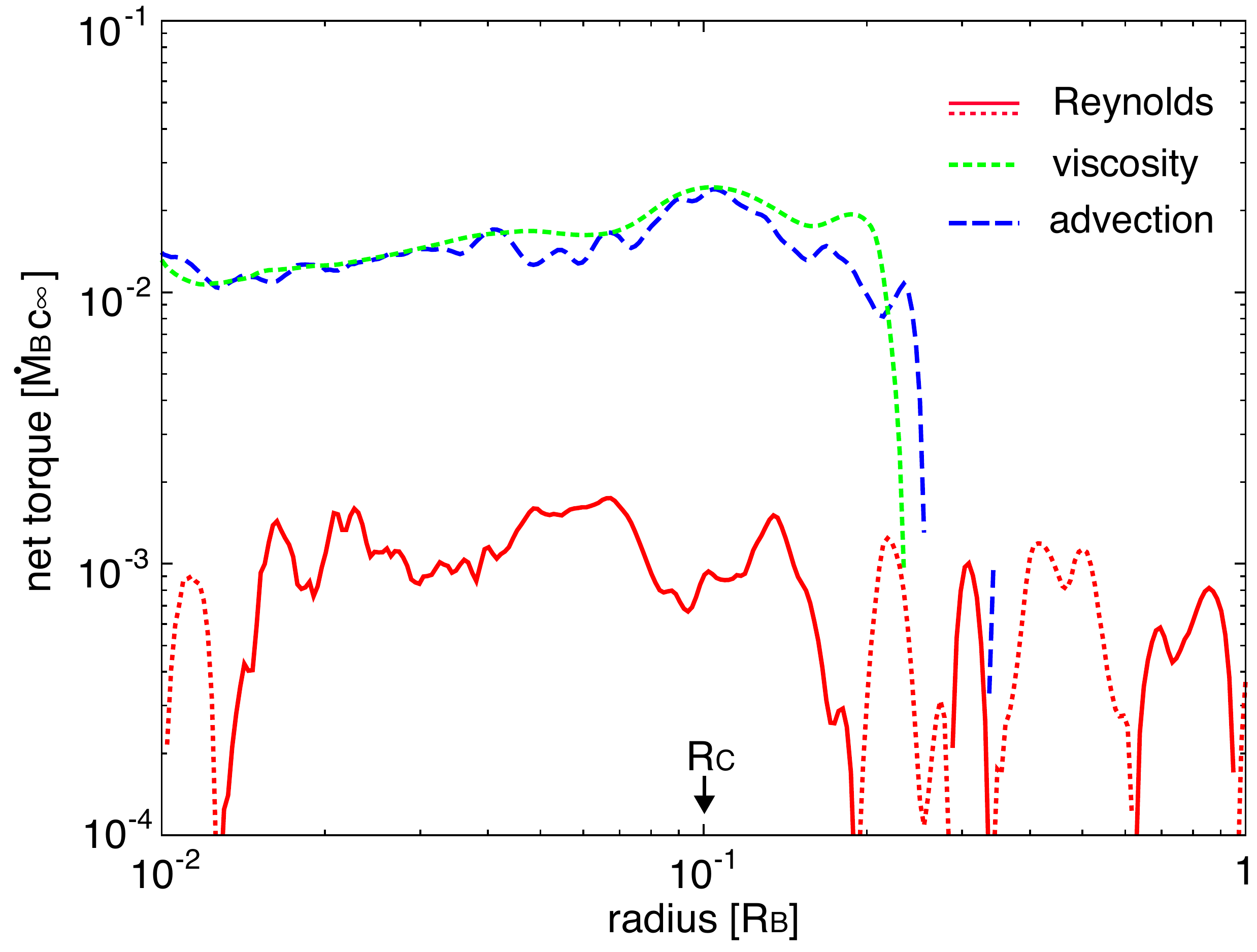}
\caption{Radial profiles of the time-averaged torque (over $6\leq t/t_{\rm dyn}\leq 16$) for the case with $R_{\rm C}/R_{\rm B}=0.1$.
Each curve shows the Reynolds torque (red), the viscous torque (green) and the advection torque (blue).
The Reynolds component due to convection transports the angular momentum inward (solid) and outward (dotted), respectively.
Within the centrifugal radius ($r\la R_{\rm C}$), the angular momentum transports via
advection (inward) and viscosity (outward) are almost balanced.
}
\label{fig:ang_flux}
\end{center}
\end{figure}

Fig.~\ref{fig:ang_flux} shows radial profiles of three kinds of torques for the case with 
$R_{\rm C}/R_{\rm B}=0.1$; the Reynolds stress (red), the viscosity (green) and the advection (blue).
Those three terms are calculated as 
\begin{equation}
T_{\rm Rey}
\equiv - \frac{\partial }{\partial r } \left \langle 2\pi r^2 \Sigma v'_r v'_\phi \right \rangle,
\end{equation}
\begin{equation}
T_{\rm vis}
\equiv \frac{\partial }{\partial r} \left\langle 2\pi r^3 \nu \Sigma \frac{\partial (v_\phi/r)}{\partial r}\right\rangle ,
\end{equation}
\begin{equation}
T_{\rm adv} \equiv -\frac{\partial }{\partial r}
\left[
\left\langle 2\pi r^2 \Sigma v_r \right\rangle 
\cdot  \left\langle v_\phi \right\rangle 
\right],
\end{equation}
where the surface density is given by $\Sigma \approx \int_0^\pi \rho r\sin^2 \theta d\theta$.
This approximation is valid when the density concentrates near the equator, 
i.e., $\rho(\theta)\propto (\sin \theta)^{2.33}$.
The sign of the torque suggests that convection motions transport angular momentum inward (red solid) 
rather than outward (red dotted), as expected from Fig.~\ref{fig:Rey}.
Within the centrifugal radius ($r\la R_{\rm C}$), the angular momentum is transported inward 
via advection ($T_{\rm adv}>0$) and outward via viscosity ($T_{\rm vis}<0$).
Both of them are tightly balanced, and the residual agrees with the Reynolds torque
within a factor of two.

\begin{figure}
\begin{center}
\includegraphics[width=82mm]{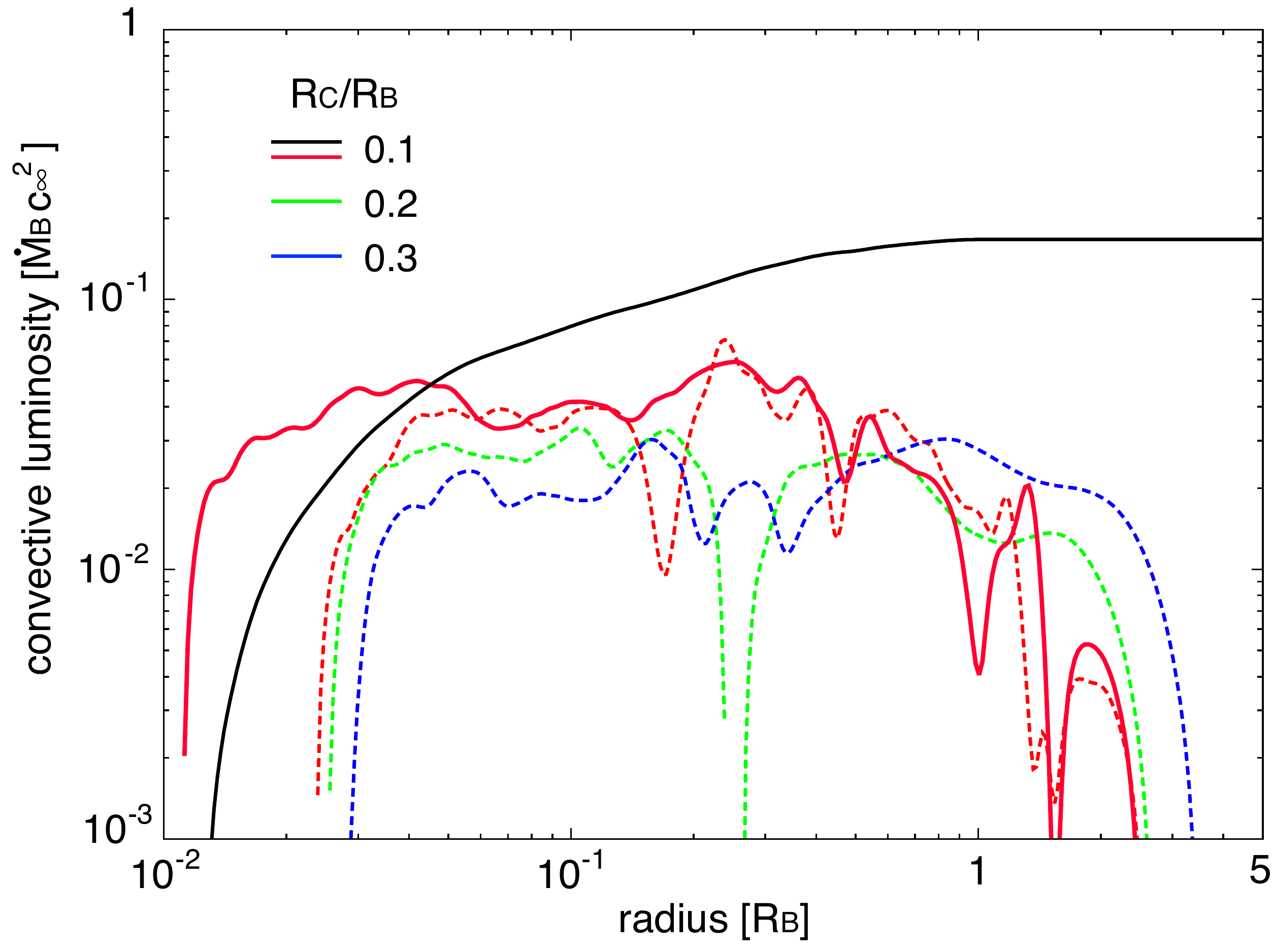}
\caption{Radial profiles of the convective luminosity, $d L_{\rm conv}/d\ln r$ (dashed) for different three cases with
$R_{\rm C}/R_{\rm B}=0.1$ (red), $0.2$ (green) and $0.3$ (blue).
In order to study the effect of the boundary condition, we conduct a simulation for $R_{\rm C}/R_{\rm B} = 0.1$ 
with a smaller $r_{\rm min}=3\times 10^{-3}~R_{\rm B}$ (red solid).
The integrated convective luminosity is shown by black solid, presenting $L_{\rm conv}\simeq 0.2~\dot{M}_{\rm B}c_\infty^2$.
}
\label{fig:L_conv}
\end{center}
\end{figure}

Convective motions can transport thermal energy as well.
In order to evaluate the convective energy flux $F_{\rm conv}$, 
we consider the time-averaged energy equation as\footnote{
The heating term due to convective viscosity and other higher-order terms are neglected.
Our purpose is not showing precise formulae but giving an order-of-magnitude 
estimate for the convective energy flux.}
\begin{equation}
\frac{1}{r^2}\frac{\partial }{\partial r}(r^2F_{\rm conv}) = 
\left\langle \sigma_{r\phi}r\frac{\partial }{\partial r}\left(\frac{v_\phi}{r}\right) \right\rangle
- \left\langle \rho Tv_r\frac{ds}{dr} \right\rangle.
\end{equation}
The two terms on the right-hand-side present time-averaged values of 
the energy generation rate by viscous dissipation 
($Q^+_{\rm vis}$) and the energy advection rate ($Q^-_{\rm adv}$), respectively.
Contribution due to the convective energy flux can be approximated by 
the difference between $\langle Q^+_{\rm vis}\rangle$ and $\langle Q^-_{\rm adv}\rangle $.
Integrating the energy equation over angles except the polar regions, we obtain 
\begin{equation}
\frac{dL_{\rm conv}}{dr}(r)=2\pi r^2 \int_{D_\theta} (\langle Q^+_{\rm vis}\rangle -\langle Q^-_{\rm adv}\rangle )\sin \theta d\theta,
\end{equation}
where the convective luminosity is defined as
$L_{\rm conv}=2\pi r^2 \int _{D_\theta}F_{\rm conv} \sin \theta d\theta$
and $D_\theta =[\pi/6, 5\pi/6]$.

\begin{figure}
\begin{center}
\includegraphics[width=82mm]{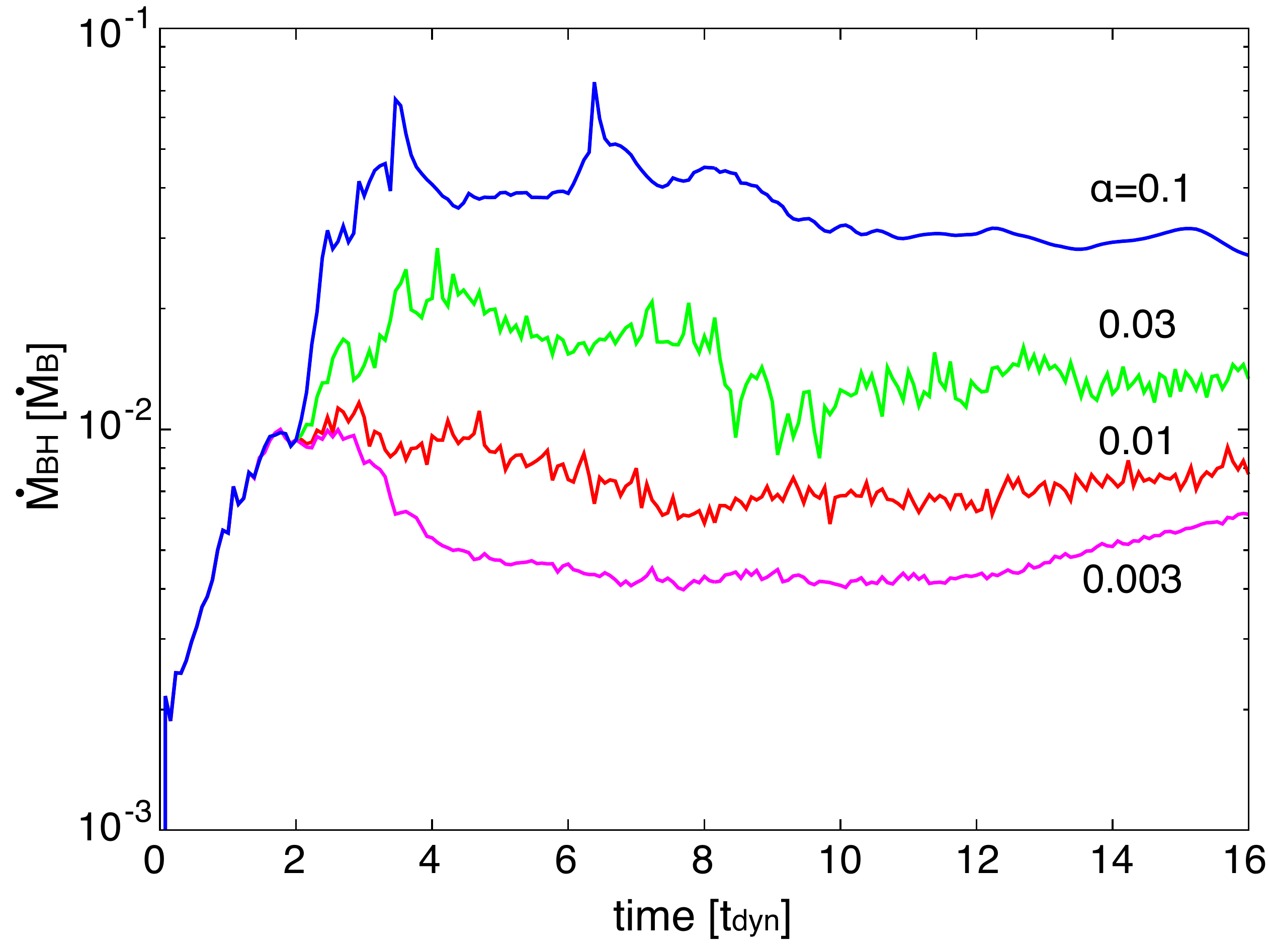}
\caption{Same as Fig.~\ref{fig:t_Mdot_j}, but 
for different viscous parameters:
$\alpha=0.003$ (magenta), $0.01$ (red), $0.03$ (green) and $0.1$ (blue).
The angular momentum is set so that $R_{\rm C}/R_{\rm B}=0.1$.
The accretion rate increases with $\alpha$
because angular momentum transport is more efficient.
}
\label{fig:t_Mdot_alpha}
\end{center}
\end{figure}

Fig. \ref{fig:L_conv} shows the radial profiles of $dL_{\rm conv}/d\ln r$ 
for different values of $R_{\rm C}/R_{\rm B}$ (dashed).
Since the derivative is positive at $0.02\la r/R_{\rm B}\la 1$, thermal energy is transported outwards by convection.
Those values are almost constant, namely $dL_{\rm conv}/d\ln r\simeq (0.02-0.04)\times \dot{M}_{\rm B}c_\infty^2$.
We note that the energy advection dominates at the inner most region ($r<0.02~R_{\rm B}$).
This is due to our inner boundary conditions, where we do not consider energy output
from the sink cell at the center.
In order to study the effect of the boundary condition, we conduct a simulation for $R_{\rm C}/R_{\rm B}=0.1$
with a smaller $r_{\rm min}$, which is set to $3\times 10^{-3}~R_{\rm B}$ (red solid).
This results clearly shows that a smaller innermost grid does not change the absolute value of $dL_{\rm conv}/d\ln r$ 
but makes the convection-dominated region larger.
We integrate the value from $10^{-2}~R_{\rm B}$ to $R_{\rm B}$ and calculate the convective luminosity (black solid).
Then, the convective luminosity is written as $L_{\rm conv}\sim ~\eta_{\rm conv}\dot{M}_{\rm B}c_\infty^2$,
where the efficiency is approximated as $\eta_{\rm conv }\simeq 0.2$.

\subsection{Effects of viscous parameter}
\label{sec:vis}

Next, we discuss the dependence of our results on the viscous parameter $\alpha$,
conducting several simulations with the same parameters as shown in \S\ref{sec:fid}
except varying $\alpha$.

Fig.~\ref{fig:t_Mdot_alpha} shows the time evolution of the accretion rate onto a BH at the center 
for different values of $\alpha$ ($0.003\leq \alpha \leq 0.1$).
To keep the numerical simulations stable, the viscous parameter is assumed to be $0.01$ at $t\leq 2~t_{\rm dyn}$,
increase (or decrease) linearly proportional to the time until $t=4~t_{\rm dyn}$, 
and keep a constant value what we consider at $t> 4~t_{\rm dyn}$.
The accretion rates increase with the viscous parameter because angular momentum transport becomes more efficient.
For the lowest value of $\alpha$, inflows with lower angular momentum through the polar region is not negligible 
because inflows driven by viscosity through the torus become less efficient.
For the highest value of $\alpha$, the fluctuation in the accretion rate is suppressed by strong viscosity,
which is consistent with that reported by previous numerical simulations by \cite{IA_1999,IA_2000}.

\begin{figure}
\begin{center}
\includegraphics[width=82mm]{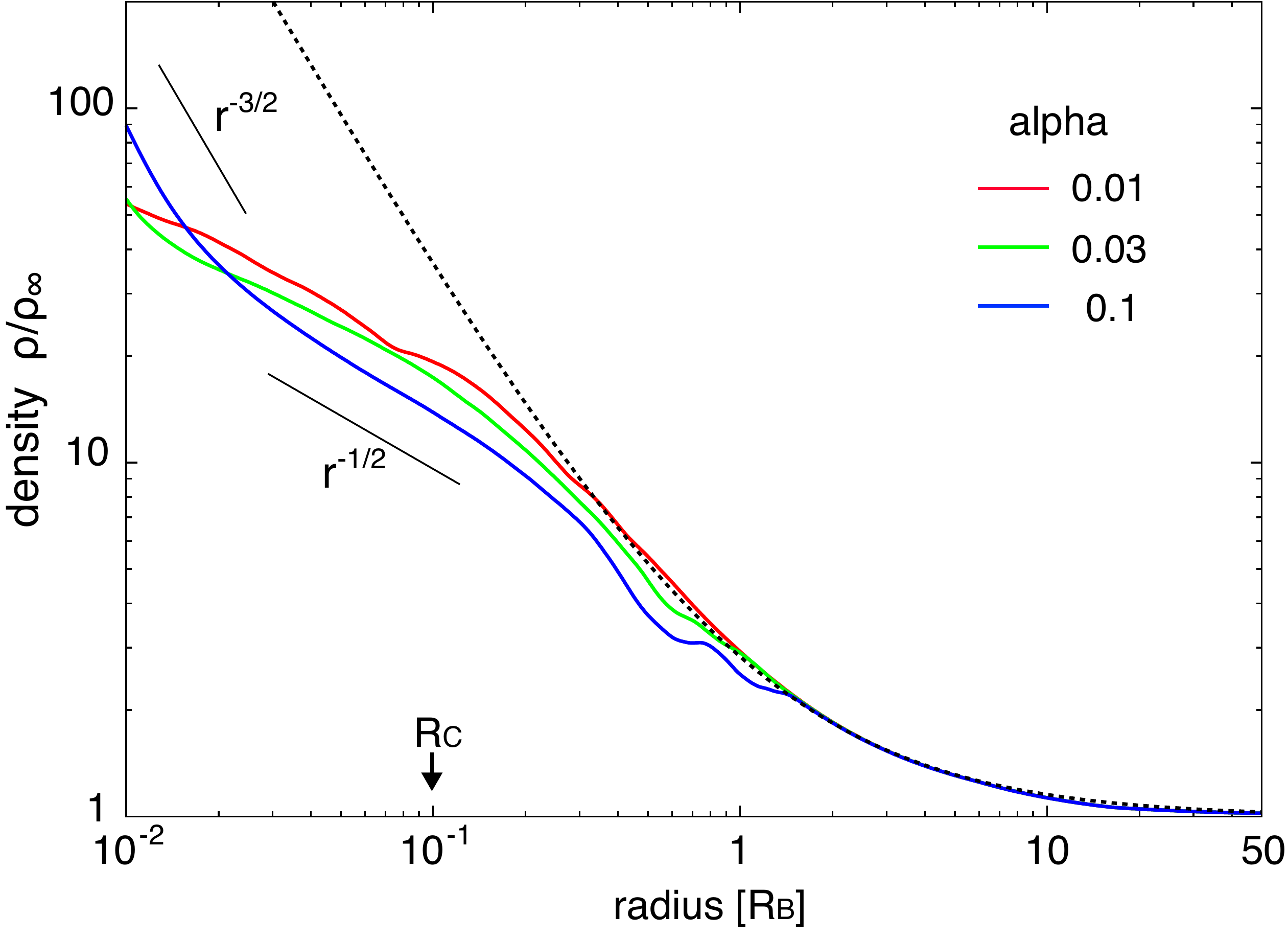}
\caption{Same as Fig.~\ref{fig:r_rho} but for different viscous parameters:
$\alpha=0.01$ (red), $0.03$ (green) and $0.1$ (blue).
The angular momentum is set to $R_{\rm C}/R_{\rm B}=0.1$.
The profiles are approximated by an equilibrium solution of $\rho /\rho_\infty = (1+R_{\rm B}/r)^{3/2}$ 
(black dotted) in the outer region and by $\rho \propto r^{-1/2}$ in the inner region ($r<0.2~R_{\rm B}$), 
which is consistent with that in CDAF solutions,
For the highest $\alpha$, the density profile approaches $\rho \propto r^{-3/2}$ in the innermost region
(see text for explanation). 
}
\label{fig:r_rho_alpha}
\end{center}
\end{figure}

\begin{figure}
\begin{center}
\includegraphics[width=82mm]{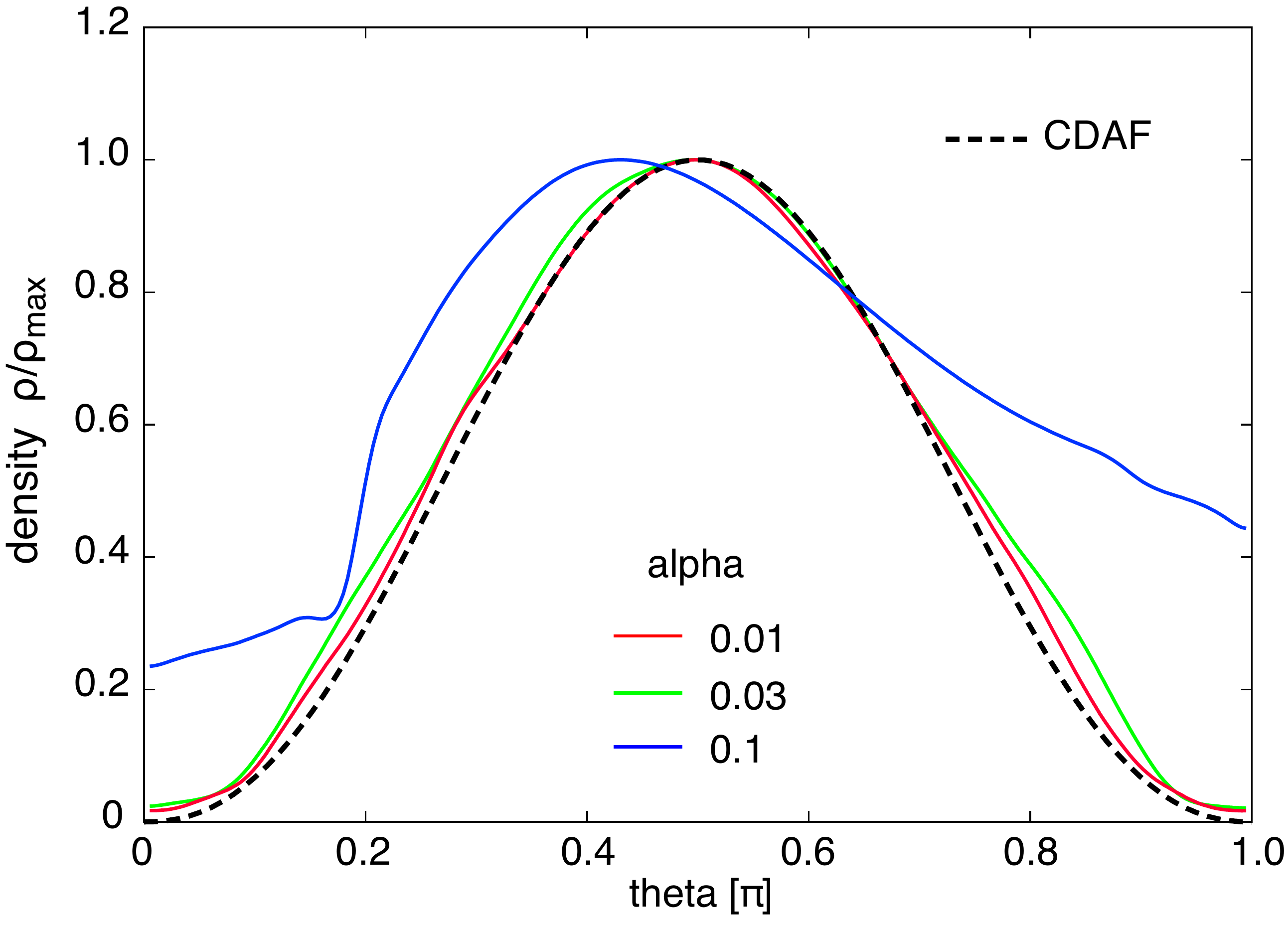}
\caption{Same as Fig.~\ref{fig:theta_rho} at $r=0.02~R_{\rm B}$ but for different viscous parameters: 
$\alpha=0.01$ (red), $0.03$ (green) and $0.1$ (blue).
The angular momentum is set to $R_{\rm C}/R_{\rm B}=0.1$.
Dashed curve presents an angular profile expected in CDAF solutions:
$\rho(\theta) \propto (\sin \theta)^{[2/(\gamma-1)-1]}$. 
}
\label{fig:theta_rho_alpha}
\end{center}
\end{figure}

In Fig.~\ref{fig:r_rho_alpha}, we show radial profiles of the gas density for three different values of the viscous parameter:
$\alpha=0.01$ (red), $0.03$ (green) and $0.1$ (blue).
The profiles are approximated by an equilibrium solution following $\rho /\rho_\infty = (1+R_{\rm B}/r)^{3/2}$ 
(dotted curve) in the outer region and $\rho \propto r^{-1/2}$, which is consistent with that in CDAF solutions
in the inner region ($r<0.2~R_{\rm B}$).
For all the cases, the overall behavior is similar, i.e., the density profile is not dependent on the choice of 
viscous parameter.
However, for the highest value of $\alpha (=0.1)$, the density slope becomes as steep as $-3/2$ at the innermost region
($r<0.02~R_{\rm B}$).
In this case, convective motions are suppressed by strong viscosity and thus the energy generated by 
viscous heating is not transported by convection but by advection with inflows, 
i.e., advection-dominated accretion flows \citep{NY_1994,NY_1995}.

Radial profiles of other physical quantities, e.g., $c_{\rm s}$, $v_{\phi}$ and $Be$, do not change 
significantly from those with $\alpha =0.01$.
However, for the highest value of $\alpha(=0.1)$, the normalized rotational velocity 
decreases at $r<0.03~R_{\rm B}$ to the center to $v_\phi/v_{\rm K}\simeq 0.3$.
Such a small ratio is found in ADAF solutions \citep{NY_1995}.

Fig.~\ref{fig:theta_rho_alpha} shows angular profiles of the gas density 
at $r=0.02~R_{\rm B}$ for different viscous parameters:
$\alpha=0.01$ (red), $0.03$ (green) and $0.1$ (blue).
For lower values of $\alpha(<0.1)$, the angular profiles of the density follow 
that of a CDAF solution (black dashed).
On the other hand, for the highest $\alpha(=0.1)$, where the radial density profile is as steep as $-3/2$, 
the angular profile is also different from the CDAF one.
Such non-equatorial symmetric profiles produced by outflows towards one of the poles 
have been reported in previous numerical simulations by \cite{IA_2000} (their Model G, where $\alpha=0.1$).

\begin{figure}
\begin{center}
\includegraphics[width=82mm]{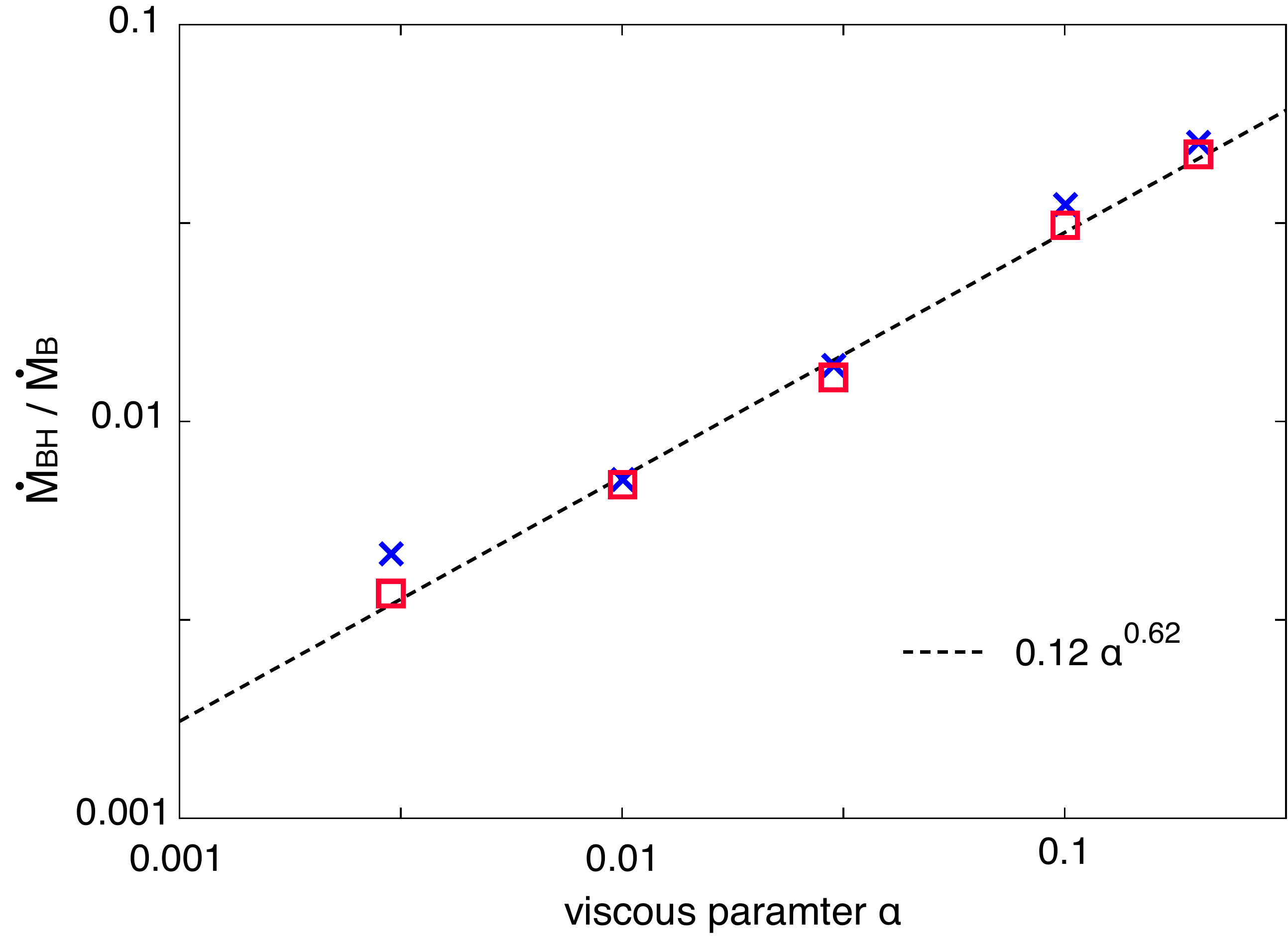}
\caption{Dependence of the net gas-inflow rate on the viscous parameter $\alpha$.
The net rate increases with the viscous parameter because angular momentum is transported more efficiently. 
The total net accretion rates (blue cross) and the inflow rate through the equatorial region (red square) are shown.
The results for $\alpha \leq 0.2$ can be fit by $\dot{M}/\dot{M}_{\rm B}\simeq 0.12~\alpha^{0.62}$ (dashed).
}
\label{fig:alpha_Mdot}
\end{center}
\end{figure}

Fig. \ref{fig:alpha_Mdot} presents the dependence of the net accretion rate on the viscous parameter $\alpha$.
Each symbol shows the total net accretion rates (blue cross) and the rate through the equatorial region 
of $\pi/6 \leq \theta \leq 5\pi/6$ (red square).
The latter can clarify the effect of angular momentum transport via viscosity in the torus.
Those results for $\alpha \leq 0.2$ can be fit by $\dot{M}/\dot{M}_{\rm B}\simeq 0.12~\alpha^\delta$,
where $\delta =0.62 \pm 0.24$ (dashed).
{\it We note that the relation is qualitatively different from that considered in previous works; self-similar solutions or 
spherical accretion solutions without treating convection}
($\dot{M}\simeq \alpha \dot{M}_{\rm B}$; \citealt{Narayan_Fabian_2011}).
Therefore, this relation between $\dot{M}$ and $\alpha$ is an essential result from 
the self-consistent solution, connecting the Fishbone-Moncrief quasi-static solution (large scales)
to existing CDAF solutions (small scales), and show the importance of the global solution because 
without our simulates, it was unclear how physical quantities on larger scales determine 
the properties of the accretion flow on small scales.

\section{Global solutions of radiatively inefficient accretion flows}
\label{sec:global}

We now briefly discuss the global solution of radiatively inefficient rotating-accretion flows,
extending our results down to the central BH.
We also give an analytical expression for the net BH feeding rate in the global solution,
which can be compared to observations of low-luminosity BHs such as Sgr A$^*$ and the BH in M87.

\subsection{BH feeding rate and energy loss via radiation}

As shown in Fig.~\ref{fig:r_Mdot}, the mass inflow rate decreases towards the center
as $\dot{M}\simeq \rho |v_r| r^2 \propto r$.
This is because the density profile of a CDAF solution follows $\rho \propto r^{-1/2}$ and 
the radial velocity follows $|v_r|\sim \nu/r \propto r^{-1/2}$ in the $\alpha$-viscosity model, respectively.
Then, the net accretion rate (including both inflows and outflows) is independent of radius 
and at a much lower rate than the Bondi value.

\begin{figure}
\begin{center}
\includegraphics[width=82mm]{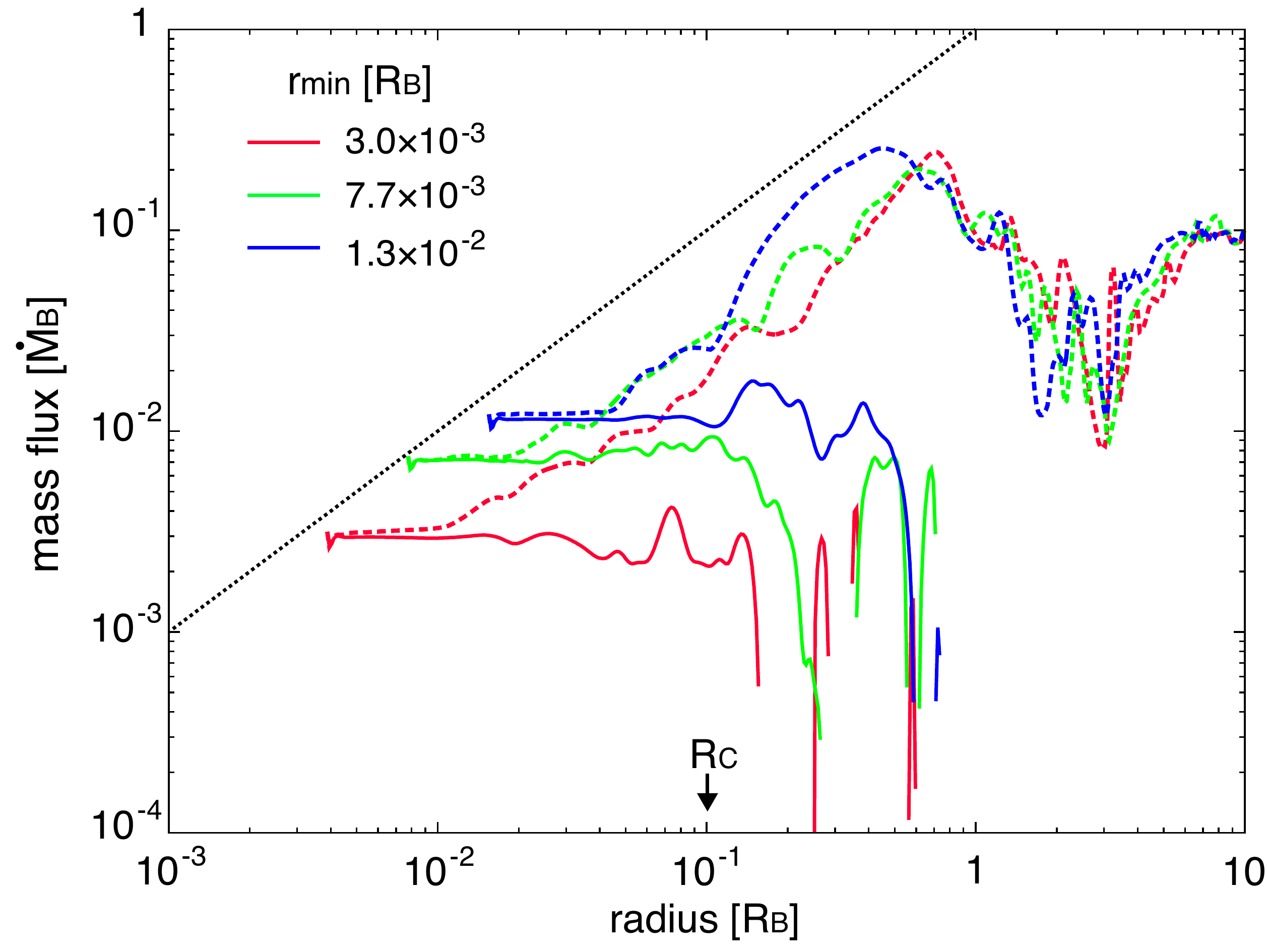}
\caption{Radial structure of the angle-integrated mass inflow and outflow rates
for different sizes of the innermost grid $r_{\rm min}$.
Physical parameters are set to $R_{\rm C}/R_{\rm B}=0.1$ and $\alpha=0.01$, and
the elapsed time is $t\simeq 13~t_{\rm dyn}$.
Solid and dashed curves present the inflow rate
and the net accretion rate (including both inflows and outflows), respectively.
Since the density follows $\rho \propto r^{-1/2}$ and the radial velocity follows $|v_r|\sim \nu/r \propto r^{-1/2}$,
the inflow rate is proportional to the radius ($\dot{M}\propto r$) within the centrifugal radius. 
The net accretion rate decreases with $r_{\rm min}$ and 
can be approximated by Eq. (\ref{eq:net_rate}).
The dotted line shows $\dot{M}=(r_{\rm min}/R_{\rm B})\dot{M}_{\rm B}$.
}
\label{fig:r_Mdot_rmin}
\end{center}
\end{figure}

Because of limitation of computational time, we do not extend our computational domain down to 
the central BH ($r\sim R_{\rm Sch}$).
Instead, we perform two additional simulations with different locations of the innermost grid 
of $r_{\rm min}=3\times 10^{-3}~R_{\rm B}$ (red) and $1.3\times 10^{-2}~R_{\rm B}$ (blue).
Fig. \ref{fig:r_Mdot_rmin} shows radial profiles of the angle-integrated inflow rate (dashed)
and the net accretion rate (solid).
As shown clearly, the net accretion rate decreases with $r_{\rm min}$.
Therefore, the time-averaged value of the net accretion rate can be approximated as 
\begin{equation}
\dot{M}\simeq \left(\frac{\alpha}{0.01}\right)^\delta 
\left(\frac{r_{\rm min}}{R_{\rm B}}\right) \dot{M}_{\rm B},
\label{eq:net_rate}
\end{equation}
where $\delta \simeq 0.62$.
Assuming that the net accretion rate is constant at $r\la 2~R_{\rm C}$,
we can estimate the net radial velocity as 
$\overline{v_r} \equiv -\dot{M}/4\pi \rho r^2$, or for $\beta (=R_{\rm C}/R_{\rm B}) \ll 1$
\begin{equation}
\overline{v_r} \simeq 
- \frac{\beta}{2}
\left(\frac{\alpha}{0.01}\right)^\delta 
\left(\frac{r_{\rm min}}{R_{\rm B}}\right)
\left(\frac{r}{R_{\rm B}}\right)^{-3/2}
~c_\infty.
\end{equation}
Using the net inflow velocity, the dynamical timescale is estimated as $t_{\rm dyn}=r/\overline{v_r}\propto r^{5/2}$.
Inside the torus, heating via viscous dissipation dominates and the heating timescale is given by
$t_{\rm vis}\simeq 1/[\gamma(\gamma-1)\alpha \Omega]^{-1}\propto r^{3/2}$.
The ratio of the two timescales is given by
\begin{equation}
\frac{t_{\rm vis}}{t_{\rm dyn}}\sim 0.6 \left(\frac{\alpha}{0.01}\right)^{\delta-1}
\left(\frac{r_{\rm min}}{10^{-2}~R_{\rm B}}\right)
\left(\frac{r}{R_{\rm C}}\right)^{-1}.
\end{equation}
Thus, we find $t_{\rm vis}\la t_{\rm dyn}$ except at $r\la 6(\beta/0.1)~r_{\rm min}$.

Next, in order to evaluate the effect of radiative cooling, we compare the heating timescale to the cooling timescale.
Since $\rho \propto r^{-1/2}$ and $T\propto r^{-1}$ inside the centrifugal radius, 
the bremsstrahlung cooling rate per volume is $Q^-_{\rm br} \propto \rho^2 T^{1/2} \propto r^{-3/2}$
and the cooling timescale is $t_{\rm cool}\propto \rho T/Q^-_{\rm br} \propto r^0$.
Since $t_{\rm vis}\la t_{\rm dyn}$ at $r\simeq R_{\rm C}$, thus we estimate the ratio of $t_{\rm vis}$ to $t_{\rm cool}$ as 
\begin{equation}
\frac{t_{\rm vis}}{t_{\rm cool}}\sim 0.25 \left(\frac{\alpha}{0.01}\right)^{-1}
\left(\frac{r}{R_{\rm C}}\right)^{3/2}
\left(\frac{T_\infty}{10^7~\K}\right)^{1/2}
\left(\frac{\dot{m}_{\rm B}}{10^{-3}}\right),
\end{equation}
where $\gamma=1.6$ and $\beta=0.1$ are assumed.
Bremsstrahlung cooling does not play an important role 
in the accretion flow as long as $\dot{m}_{\rm B}<4\times 10^{-3}$,
which is consistent with the results shown in \cite{LOS_2013}.
Since $Q^-_{\rm br} \propto r^{-3/2}$, the convection-dominated accretion flow 
is sandwiched between an inner region where gravitational energy would be released 
close to the BH and an outer radiating region \citep{Quataert_2000,Narayan_2000,Ball_2001}.

\subsection{Thermal conductivity}
\label{sec:cond}

In a hot accretion flow, thermal conductivity potentially transports energy outward 
and could affect the gas dynamics \citep[e.g.,][]{Johnson_Quataert_2007,Sharma_2008,Shcherbakov_2010}.
In the classic picture, the conductive energy flux is estimated as 
\begin{equation}
\mbox{\boldmath $F$}_{\rm cond}=-\kappa \mbox{\boldmath $\nabla$}T,
\end{equation}
where $\kappa$ is the conduction coefficient given by \cite{Spitzer_1962} as
\begin{equation}
\kappa = 
5.0\times 10^{-7} \left(\frac{\ln \Lambda_{\rm c}}{37}\right)^{-1}T^{5/2} f_c,
\end{equation}
in units of ${\rm erg~s^{-1}~cm^{-1}~K^{-1}}$, and $f_c$ is the conductivity suppression factor 
because thermal conduction in the perpendicular directions to magnetic fields can be suppressed.
The value of the suppression factor has been discussed by various theoretical arguments and estimated 
as $f_c\sim 0.1$ \citep[e.g., ][]{Narayan_Medvedev_2001, Maron_2004}.
However, the precise value of the suppression factor is highly uncertain and 
depends on the nature of magnetic turbulence.
It is known that there are two instabilities of magnetized plasma:
the magneto-thermal instability (MTI; \citealt{Balbus_2000}) 
and the heat-flux-driven buoyancy instability (HBI; \citealt{Quataert_2008}).
The MTI occurs when the temperature decreases with height and the HBI does in the opposite case.
\cite{McCourt_2011} have performed long-term and global MHD simulations in order to 
follow the evolution of both instabilities into the non-linear regime.
They have found the following two results:
(1) the MTI can drive strong turbulence leading to the isotropic configuration of magnetic fields ($f_c\sim 0.3$) 
and produce a large convective energy flux of $\ga 0.01 \rho c_s^3$, and 
(2) The HBI reorients the magnetic field and suppresses the conductive heat flux through the plasma.
Since the value of the suppression factor is uncertain, we adopt $f_c=0.1$ as a fiducial one.

Using our accretion solution inside the centrifugal radius, where 
the temperature follows $T\simeq T_\infty f(\gamma)R_{\rm B}/r$,
the conductive luminosity ($L_{\rm cond} \equiv 4\pi r^2 F_{\rm cond}$)
is estimated as 
\begin{equation}
L_{\rm cond} \simeq 
6.4\times 10^{29}~M_6 c_7^{-2}
\left(\frac{f_c}{0.1}\right)
\left(\frac{r}{R_{\rm B}}\right)^{-5/2}
~{\rm erg~s^{-1}}.
\end{equation}
On the other hand, the convective luminosity is roughly given by 
\begin{align}
L_{\rm conv} &\simeq \eta_{\rm conv} \left(\frac{\alpha}{0.01}\right)^\delta
\dot{M}_{\rm B} c_\infty^2,\nonumber\\
&\simeq 1.1\times 10^{35}~\rho_{-22}M_6^2 c_7^{-1}
\left(\frac{\eta_{\rm conv}}{0.2}\right)
\left(\frac{\alpha}{0.01}\right)^\delta
~{\rm erg~s^{-1}}.
\end{align}
Thus, the ratio of the two luminosities is 
\begin{align}
\frac{L_{\rm cond}}{L_{\rm conv}} 
&\simeq 4.5\times 10^{-8}
\left(\frac{f_c}{0.1}\right)
\left(\frac{\eta_{\rm conv}}{0.2}\right)^{-1}
\left(\frac{\alpha}{0.01}\right)^{-\delta}
\nonumber\\
&~~~\times 
\left(\frac{\dot{m}_{\rm B}}{10^{-3}}\right)^{-1}
\left(\frac{T_\infty}{10^7~\K}\right)^{-2}
\left(\frac{r}{R_{\rm B}}\right)^{-5/2}.
\end{align}
Therefore, we can estimate a characteristic radius where $L_{\rm conv}\simeq L_{\rm cond}$ as 
\begin{align}
\frac{R_{\rm tr}}{R_{\rm B}} & \simeq 1.5\times 10^{-3}
\left(\frac{f_c}{\eta_{\rm conv}}\right)^{2/5}
\left(\frac{\alpha}{0.01}\right)^{-2\delta/5}
\nonumber\\
& ~~~ \times \left(\frac{\dot{m}_{\rm B}}{10^{-3}}\right)^{-2/5}
\left(\frac{T_\infty}{10^7~\K}\right)^{-4/5}.
\label{eq:r_tr}
\end{align}
Inside this radius, thermal conduction would dominate the energy transport over convection,
and the temperature profile would be flatter rather than the adiabatic scaling law ($T\propto r^{-1}$).
The ratio of $c_{\rm s}/v_{\rm K}$ would decline and the flow would resemble an optically thin 
viscous disk.
Note that thermal conduction would hardly affect the dynamics of accretion flows at the vicinity of the BH 
($r<100~R_{\rm Sch}$), where the conductive energy flux is limited to a few percent of the saturated value 
of $\sim \rho c_{\rm s}^3$ \citep{Foucart_2016,Foucart_2017}.
This fraction is consistent with that obtained in our solution at $r\ga R_{\rm tr}$, namely 
$L_{\rm cond}/(4\pi r^2 \rho c_{\rm s}^3)\sim 1.5\times 10^{-2}(r/R_{\rm tr})^{-5/2}$.
To explore the nature of thermal conduction on the convective motions is left for future investigations.

\subsection{Connection to the inner-most region}

As discussed in \S\ref{sec:cond}, some physical processes, e.g., thermal conduction, 
can transport the energy outward instead of convection.
Once additional energy transport operates, {\it if any}, the temperature profile is no longer 
that for an adiabatic flow.
We characterize the transition radius as $R_{\rm tr} =\xi R_{\rm B}$ ($\xi\ll1$).
The critical radius where $L_{\rm conv}\simeq L_{\rm cond}$ corresponds to $\xi =\xi_{\rm cond}(\simeq 10^{-3})$.
Since the temperature follows $T\propto r^{-p}$ ($0\la p< 1$) within $R_{\rm tr}$, 
the accretion disk becomes thinner and thus the density follows $\rho \propto r^{-3+3p/2}$,
assuming that the accretion rate is constant through the disk as 
$\dot{M}\simeq \xi  (\alpha/0.01)^\delta \dot{M}_{\rm B}$ (see Eq. \ref{eq:net_rate}).
Inside the transition radius, the optical depth to electron scattering increases and 
is estimated as
\begin{align}
\tau_{\rm es}(r)&=\int_r^{R_{\rm tr}} \rho \kappa_{\rm es}dr,\nonumber \\
&\simeq \frac{40\xi^{1/2}}{4-3p}
\frac{\dot{m}_{\rm B}}{\beta}
\left(\frac{c_\infty}{c}\right)
\left(\frac{r}{R_{\rm tr}}\right)^{-2+3p/2},
\end{align}
where $\beta\ll 1$ is assumed.
We evaluate the optical depth at the innermost stable circular orbit 
(ISCO; $r\sim 3~R_{\rm Sch}$) for $p=0$ as
\begin{align}
\tau_{\rm es}^{\rm ISCO}
\simeq 0.022
\left(\frac{\beta}{0.1}\right)^{-1}
\left(\frac{\xi}{\xi_{\rm cond}}\right)^{5/2}
\left(\frac{\dot{m}_{\rm B}}{10^{-3}}\right)
\left(\frac{T_\infty}{10^7~\K}\right)^{3/2}.
\end{align}
Since the accretion flow at the ISCO is optically thin even in an isothermal case ($p=0$), 
the gas is more likely to be optically thin for $0<p<1$.
In this paper, our numerical simulations do not treat thermal conductivity because 
the effect is subdominant in the computation domain of $r\geq r_{\rm min}(>R_{\rm tr})$.
We plan to numerically study the transition to the innermost solution in future work.

\subsection{Radiation luminosities of BHs with very low accretion rates}

Finally, we estimate radiation luminosities produced from a low-density, radiatively inefficient accretion flow.
Although our simulations do not treat the inner region ($r\la R_{\rm tr}$), where energy transport by 
thermal conduction would dominate that by convection,
we can infer the accretion rate onto the central BH from Eqs. (\ref{eq:net_rate}) and (\ref{eq:r_tr}),
\begin{align}
\frac{\dot{M}}{\dot{M}_{\rm Edd}}\simeq 1.5\times 10^{-6}
\left(\frac{\alpha}{0.01}\right)^{3\delta/5} 
\left(\frac{T_\infty}{10^7~\K}\right)^{-4/5}
\left(\frac{\dot{m}_{\rm B}}{10^{-3}}\right)^{3/5},
\label{eq:net_rate_BH}
\end{align}
where $f_c/\eta_{\rm conv}=1$ is set.
In order to estimate the radiation luminosity, we adopt a radiative-efficiency model for 
different values of $\dot{M}/\dot{M}_{\rm Edd}$, provided by axisymmetric numerical 
simulations of accretion flows around a supermassive BH with $M_\bullet=10^8~\msun$ 
on a small scale ($r\leq 100~R_{\rm Sch}$), 
including general relativistic MHD and frequency-dependent radiation transport \citep{Ryan_2017}
(see also \citealt{Ohsuga_2009,Sadowski_2017}).
We note that their simulation results including radiation processes and electron-thermodynamics,
are no longer scale-free, i.e., the radiative efficiency we adopt probably depend on the choice of the BH mass.
Nevertheless, it is worth demonstrating the radiation luminosity produced from a self-consistent, radiatively inefficient 
accretion flow, assuming that the radiative efficiency hardly depends on the BH mass.

\begin{figure}
\begin{center}
\includegraphics[width=82mm]{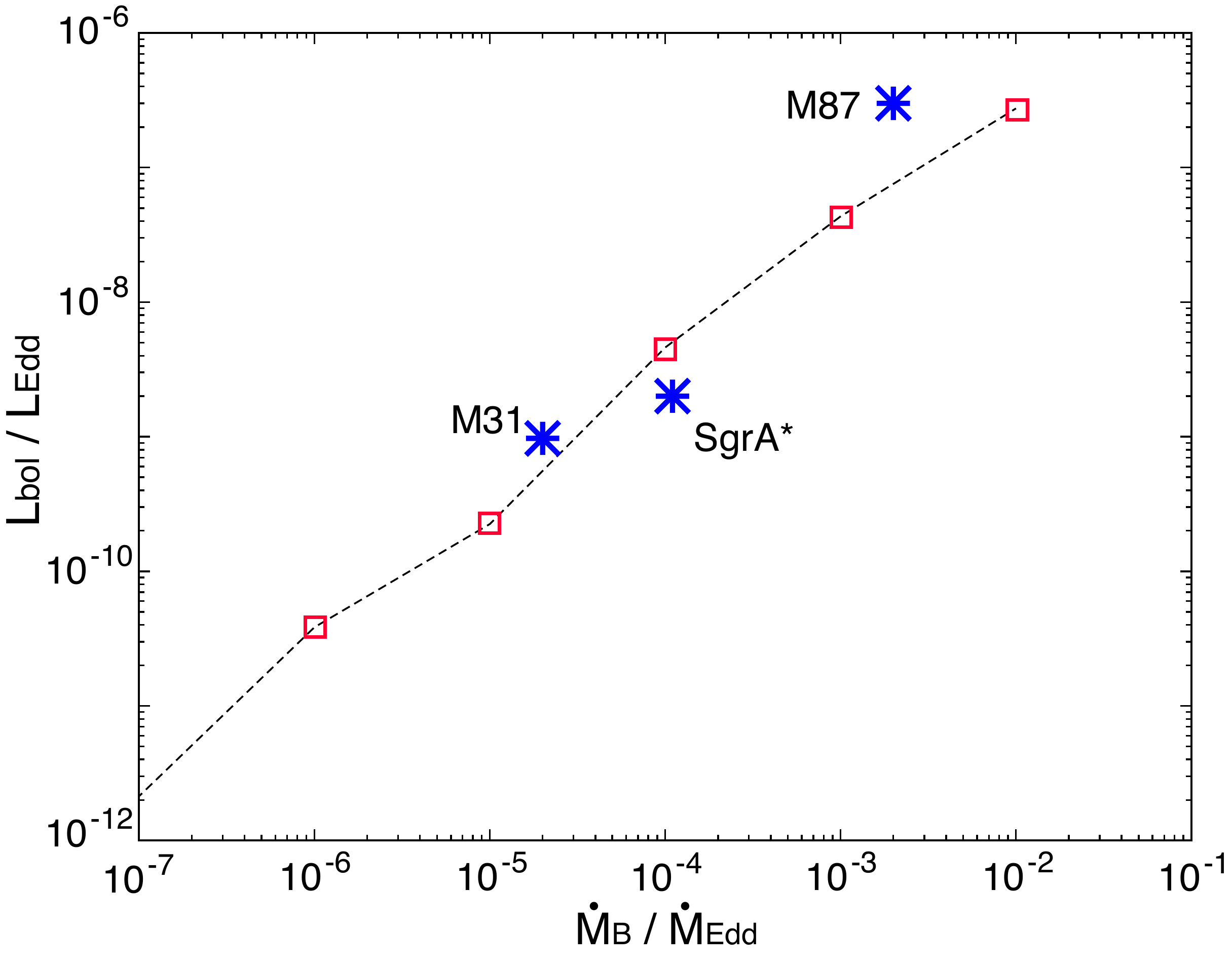}
\caption{Bolometric radiation luminosity produced by radiatively inefficient accretion flows (red square).
The luminosities are calculated from our results given by  Eqs. (\ref{eq:net_rate}) and (\ref{eq:r_tr}),
combining with a radiative efficiency provided by MHD simulations on a small scale 
($r\leq 100~R_{\rm Sch}$), including general relativity and radiation transfer \citep{Ryan_2017}.
We also present observational results for SgrA$^*$ and BHs in M31 and M87 (blue asterisk).}
\label{fig:L_rad}
\end{center}
\end{figure}

Fig~\ref{fig:L_rad} shows the radiation luminosities estimated from our results (red square)
as a function of the Bondi rate normalized by the Eddington luminosity.
The radiation luminosity produced in the vicinity of the BH is extremely low
because (1) the radiative efficiency decreases with the accretion rate in such low-accretion-rate regime 
($L\propto \dot{M}^{1.7}$; see Fig. 1 in \citealt{Ryan_2017}) and (2) the net accretion rate at $r\la R_{\rm C}$ is strongly suppressed by 
convective motions from the Bondi accretion rate $\dot{M}_{\rm B}$.
We can approximate the radiation luminosity for $\dot{M}/\dot{M}_{\rm Edd}\la 10^{-3}$ as 
\begin{equation}
\frac{L_{\rm bol}}{L_{\rm Edd}} \simeq 3\times 10^{-5}
\left(\frac{\alpha}{0.01}\right)^\delta 
\frac{\dot{M}_{\rm B}}{\dot{M}_{\rm Edd}},
\end{equation}
or
\begin{equation}
L_{\rm bol} \simeq 3\times 10^{-6} \dot{M}_{\rm B}c^2 \left(\frac{\alpha}{0.01}\right)^\delta.
\label{eq:Lbol}
\end{equation}
We also present observational results of SgrA$^*$ and BHs in M31 and M87 in 
(blue symbols in Fig~\ref{fig:L_rad}).
Those supermassive BHs in the nearby Universe are known as accreting BHs 
at low rates of $\dot{M}_{\rm B}/\dot{M}_{\rm Edd}\sim 10^{-5}-10^{-3}$.
Moreover, gas properties in the nuclei are studied since the angular sizes of their Bondi radii ($\ga 1"$) 
can be resolved well by {\it Chandra} observations
\citep[e.g., ][and references therein]{Baganoff_2003, Russell_2015,Garcia_2010}\footnote{
The bolometric luminosity of the M31 BH is estimated from the X-ray luminosity, 
assuming a bolometric correction factor of $f_{\rm bol}=10$ for such 
a low-luminosity BH \citep{Hopkins_2007}.}.
Our theoretical estimate agrees well with the observational results for SgrA$^*$ and M31.
Although the observed luminosity of M87 is several times higher than our estimate,
this would be because suppression of the accretion by convective motions in the CDAF regime
becomes less efficient for $\dot{M}_{\rm B}/\dot{M}_{\rm Edd}\ga 10^{-3}$, 
and/or because there is a level of contamination from the dominant emission due to the jet.

Isolated BHs in our Galaxy can also accrete interstellar medium (ISM) at very low rates.
Assuming typical density of the ISM and molecular clouds
($n\sim 1-100~\cc$) and the BH has a peculiar velocity of $V\sim 40-100~\kms$,
the Bondi-Hoyle-Lyttleton accretion rate is as low as 
$\dot{M}_{\rm B}/\dot{M}_{\rm Edd}\sim 10^{-8}-10^{-5}$
for $M_\bullet \sim 10~\msun$.
Some of those isolated BHs in our Galaxy could be observed as high-energy sources
\citep{Fujita_1998,Armitage_Natarajan_1999,Agol_2002,Ioka_2017,Matsumoto_2017}.
However, since those previous studies assumed self-similar ADAF solutions, 
it is worth revisiting their arguments considering our self-consistent accretion solutions.
Note that \cite{Perna_2003} have discussed the same problem for neutron stars, and 
pointed out that the reduction of accretion rates from the Bondi rates can explain why 
isolated neutron stars accreting the ISM are rarely observed in X-rays.

\section{Summary}
\label{sec:sum}

We study radiatively inefficient rotating accretion flows onto a black hole (BH) 
with two-dimensional hydrodynamical simulations.
We consider axisymmetric accretion flows and allow angular momentum transport adopting 
the $\alpha$ viscosity prescription.
When the gas angular momentum is low enough to form a rotationally supported disk/torus
within the Bondi radius ($R_{\rm B}$), we find a global steady accretion solution.
The solution consists of three phases: (1) a rotational equilibrium distribution at $r\sim R_{\rm B}$,
where the density distribution follows $\rho \propto (1+R_{\rm B}/r)^{3/2}$ and the gas is very subsonic;
(2) a geometrically thick accretion torus at the centrifugal radius $R_{\rm C}(<R_{\rm B})$, 
where thermal energy generated by viscosity is transported via strong convection motions 
expected in convection-dominated accretion flows (CDAF);
and (3) an estimated inner optically-thin disk controlled by conduction and viscosity.
Physical properties of the accretion flow in the intermediate region (2) agree with those in CDAFs
(e.g., $\rho \propto r^{-1/2}$).
In the CDAF solution, the gas accretion rate decreases towards the center due to convection ($\dot{M}\propto r$),
and the net accretion rate is strongly suppressed by several orders of magnitude from the Bondi accretion rate.
We find that the net accretion rate depends on the viscous strength and the size of the innermost radius of the 
computational domain ($r_{\rm min}$), and can be approximated as 
$\dot{M}\sim (\alpha/0.01)^{0.6}(r_{\rm min}/R_{\rm B})\dot{M}_{\rm B}$.
This solution holds for low accretion rates of $\dot{M}_{\rm B}/\dot{M}_{\rm Edd}\la 10^{-3}$
having minimal radiation cooling.
In a hot plasma at the bottom of this solution ($r<10^{-3}~R_{\rm B}\la r_{\rm min}$), 
thermal conduction would dominate the convective energy flux.
Since suppression of the accretion by convection ceases, the final BH feeding rate is found to be 
$\dot{M}/\dot{M}_{\rm B} \sim 10^{-2}-10^{-3}$.
This rate is as low as $\dot{M}/\dot{M}_{\rm Edd} \sim 10^{-7}-10^{-6}$ 
inferred for SgrA$^*$ and the nuclear BHs in M31 and M87, and 
can explain the low accretion rates and low luminosities in these sources, without invoking any feedback mechanism.

Finally, we briefly mention several effects which we do not treat in our simulations.
First, stellar winds and X-ray emission from massive stars around the Bondi radius are neglected.
These effects heat the gas to $T\sim {\rm keV}$ and could drive outflows as in 
the Galactic center around SgrA$^*$ \citep[e.g.,][]{Najarro_1997, Baganoff_2003,Quataert_2004}.
Secondly, non-axisymmetric perturbations would lead to instability and allow additional angular momentum 
transport in the accretion flow \citep[e.g.,][]{Papaloizou_Pringle_1984, Chandrasekhar_1961}.
In order to include these effects, we further need to conduct three-dimensional hydrodynamical 
simulations of accretion flows at large scales \citep[e.g.,][]{Gaspari_2013}.
Thirdly, thermal conductivity in the inner region ($r<R_{\rm tr}\sim 10^{-3}~R_{\rm B}$) 
plays an important role as an efficient process carrying the energy outwards instead of convection.
In fact, the location of the transition radius would determine the net accretion rate onto the nuclear disk
and the radiative luminosity produced in the vicinity of the BH as shown in Fig.~\ref{fig:L_rad}.
We need to investigate the detailed properties of radiatively inefficient accretion flows, 
in order to make a prediction for future observations with 
the Event Horizon Telescope\footnote{http://eventhorizontelescope.org}.

\section*{Acknowledgements}
We thank James Stone, Charles Gammie, Ramesh Narayan, Eliot Quataert, Lorenzo Sironi, Daniel Wang, 
Kengo Tomida, Kazumi Kashiyama, Kohei Ichikawa, Takashi Hosokawa and Kazuyuki Sugimura for useful discussions.
This work is partially supported by the Simons Foundation through the Simons Society of Fellows (KI),
by a Simons Fellowship in Theoretical Physics (ZH), and by NASA grant NNX15AB19G (ZH).
RK acknowledges financial support via the Emmy Noether Research Group on Accretion Flows and 
Feedback in Realistic Models of Massive Star Formation funded by the German Research Foundation 
(DFG) under grant no. KU 2849/3-1.
Numerical computations were carried out on Cray XC30 at the Center for Computational Astrophysics 
of the National Astronomical Observatory of Japan.

\bibliographystyle{mnras}
{\small
\bibliography{ref}

\begin{thebibliography}{}
\makeatletter
\relax
\def\mn@urlcharsother{\let\do\@makeother \do\$\do\&\do\#\do\^\do\_\do\%\do\~}
\def\mn@doi{\begingroup\mn@urlcharsother \@ifnextchar [ {\mn@doi@}
  {\mn@doi@[]}}
\def\mn@doi@[#1]#2{\def\@tempa{#1}\ifx\@tempa\@empty \href
  {http://dx.doi.org/#2} {doi:#2}\else \href {http://dx.doi.org/#2} {#1}\fi
  \endgroup}
\def\mn@eprint#1#2{\mn@eprint@#1:#2::\@nil}
\def\mn@eprint@arXiv#1{\href {http://arxiv.org/abs/#1} {{\tt arXiv:#1}}}
\def\mn@eprint@dblp#1{\href {http://dblp.uni-trier.de/rec/bibtex/#1.xml}
  {dblp:#1}}
\def\mn@eprint@#1:#2:#3:#4\@nil{\def\@tempa {#1}\def\@tempb {#2}\def\@tempc
  {#3}\ifx \@tempc \@empty \let \@tempc \@tempb \let \@tempb \@tempa \fi \ifx
  \@tempb \@empty \def\@tempb {arXiv}\fi \@ifundefined
  {mn@eprint@\@tempb}{\@tempb:\@tempc}{\expandafter \expandafter \csname
  mn@eprint@\@tempb\endcsname \expandafter{\@tempc}}}

\bibitem[\protect\citeauthoryear{{Agol} \& {Kamionkowski}}{{Agol} \&
  {Kamionkowski}}{2002}]{Agol_2002}
{Agol} E.,  {Kamionkowski} M.,  2002, \mn@doi [\mnras]
  {10.1046/j.1365-8711.2002.05523.x}, \href
  {http://adsabs.harvard.edu/abs/2002MNRAS.334..553A} {334, 553}

\bibitem[\protect\citeauthoryear{{Armitage} \& {Natarajan}}{{Armitage} \&
  {Natarajan}}{1999}]{Armitage_Natarajan_1999}
{Armitage} P.~J.,  {Natarajan} P.,  1999, \mn@doi [\apjl] {10.1086/312261},
  \href {http://adsabs.harvard.edu/abs/1999ApJ...523L...7A} {523, L7}

\bibitem[\protect\citeauthoryear{{Baganoff} et~al.,}{{Baganoff}
  et~al.}{2003}]{Baganoff_2003}
{Baganoff} F.~K.,  et~al., 2003, \mn@doi [\apj] {10.1086/375145}, \href
  {http://adsabs.harvard.edu/abs/2003ApJ...591..891B} {591, 891}

\bibitem[\protect\citeauthoryear{{Balbus}}{{Balbus}}{2000}]{Balbus_2000}
{Balbus} S.~A.,  2000, \mn@doi [\apj] {10.1086/308732}, \href
  {http://adsabs.harvard.edu/abs/2000ApJ...534..420B} {534, 420}

\bibitem[\protect\citeauthoryear{{Balbus} \& {Hawley}}{{Balbus} \&
  {Hawley}}{1991}]{Balbus_Hawley_1991}
{Balbus} S.~A.,  {Hawley} J.~F.,  1991, \mn@doi [\apj] {10.1086/170270}, \href
  {http://adsabs.harvard.edu/abs/1991ApJ...376..214B} {376, 214}

\bibitem[\protect\citeauthoryear{{Balbus} \& {Hawley}}{{Balbus} \&
  {Hawley}}{1998}]{Balbus_Hawley_1998}
{Balbus} S.~A.,  {Hawley} J.~F.,  1998, \mn@doi [Reviews of Modern Physics]
  {10.1103/RevModPhys.70.1}, \href
  {http://adsabs.harvard.edu/abs/1998RvMP...70....1B} {70, 1}

\bibitem[\protect\citeauthoryear{{Ball}, {Narayan}  \& {Quataert}}{{Ball}
  et~al.}{2001}]{Ball_2001}
{Ball} G.~H.,  {Narayan} R.,   {Quataert} E.,  2001, \mn@doi [\apj]
  {10.1086/320465}, \href {http://adsabs.harvard.edu/abs/2001ApJ...552..221B}
  {552, 221}

\bibitem[\protect\citeauthoryear{{Bender} et~al.,}{{Bender}
  et~al.}{2005}]{Bender_2005}
{Bender} R.,  et~al., 2005, \mn@doi [\apj] {10.1086/432434}, \href
  {http://adsabs.harvard.edu/abs/2005ApJ...631..280B} {631, 280}

\bibitem[\protect\citeauthoryear{{Blandford} \& {Begelman}}{{Blandford} \&
  {Begelman}}{1999}]{Blandford_Begelman_1999}
{Blandford} R.~D.,  {Begelman} M.~C.,  1999, \mn@doi [\mnras]
  {10.1046/j.1365-8711.1999.02358.x}, \href
  {http://adsabs.harvard.edu/abs/1999MNRAS.303L...1B} {303, L1}

\bibitem[\protect\citeauthoryear{{Blandford} \& {Begelman}}{{Blandford} \&
  {Begelman}}{2004}]{Blandford_Begelman_2004}
{Blandford} R.~D.,  {Begelman} M.~C.,  2004, \mn@doi [\mnras]
  {10.1111/j.1365-2966.2004.07425.x}, \href
  {http://adsabs.harvard.edu/abs/2004MNRAS.349...68B} {349, 68}

\bibitem[\protect\citeauthoryear{{Bondi}}{{Bondi}}{1952}]{Bondi_1952}
{Bondi} H.,  1952, \mnras, \href
  {http://adsabs.harvard.edu/abs/1952MNRAS.112..195B} {112, 195}

\bibitem[\protect\citeauthoryear{{Chandrasekhar}}{{Chandrasekhar}}{1961}]{Chandrasekhar_1961}
{Chandrasekhar} S.,  1961, {Hydrodynamic and hydromagnetic stability}

\bibitem[\protect\citeauthoryear{{Ciotti} \& {Ostriker}}{{Ciotti} \&
  {Ostriker}}{2001}]{Ciotti_Ostriker_2001}
{Ciotti} L.,  {Ostriker} J.~P.,  2001, \mn@doi [\apj] {10.1086/320053}, \href
  {http://adsabs.harvard.edu/abs/2001ApJ...551..131C} {551, 131}

\bibitem[\protect\citeauthoryear{{Ciotti}, {Ostriker}  \& {Proga}}{{Ciotti}
  et~al.}{2009}]{Ciotti_2009}
{Ciotti} L.,  {Ostriker} J.~P.,   {Proga} D.,  2009, \mn@doi [\apj]
  {10.1088/0004-637X/699/1/89}, \href
  {http://adsabs.harvard.edu/abs/2009ApJ...699...89C} {699, 89}

\bibitem[\protect\citeauthoryear{{Fishbone} \& {Moncrief}}{{Fishbone} \&
  {Moncrief}}{1976}]{Fishbone_Moncrief_1976}
{Fishbone} L.~G.,  {Moncrief} V.,  1976, \mn@doi [\apj] {10.1086/154565}, \href
  {http://adsabs.harvard.edu/abs/1976ApJ...207..962F} {207, 962}

\bibitem[\protect\citeauthoryear{{Foucart}, {Chandra}, {Gammie}  \&
  {Quataert}}{{Foucart} et~al.}{2016}]{Foucart_2016}
{Foucart} F.,  {Chandra} M.,  {Gammie} C.~F.,   {Quataert} E.,  2016, \mn@doi
  [\mnras] {10.1093/mnras/stv2687}, \href
  {http://cdsads.u-strasbg.fr/abs/2016MNRAS.456.1332F} {456, 1332}

\bibitem[\protect\citeauthoryear{{Foucart}, {Chandra}, {Gammie}, {Quataert}  \&
  {Tchekhovskoy}}{{Foucart} et~al.}{2017}]{Foucart_2017}
{Foucart} F.,  {Chandra} M.,  {Gammie} C.~F.,  {Quataert} E.,   {Tchekhovskoy}
  A.,  2017, \mn@doi [\mnras] {10.1093/mnras/stx1368}, \href
  {http://cdsads.u-strasbg.fr/abs/2017MNRAS.470.2240F} {470, 2240}

\bibitem[\protect\citeauthoryear{{Fujita}, {Inoue}, {Nakamura}, {Manmoto}  \&
  {Nakamura}}{{Fujita} et~al.}{1998}]{Fujita_1998}
{Fujita} Y.,  {Inoue} S.,  {Nakamura} T.,  {Manmoto} T.,   {Nakamura} K.~E.,
  1998, \mn@doi [\apjl] {10.1086/311220}, \href
  {http://adsabs.harvard.edu/abs/1998ApJ...495L..85F} {495, L85}

\bibitem[\protect\citeauthoryear{{Garcia} et~al.,}{{Garcia}
  et~al.}{2010}]{Garcia_2010}
{Garcia} M.~R.,  et~al., 2010, \mn@doi [\apj] {10.1088/0004-637X/710/1/755},
  \href {http://adsabs.harvard.edu/abs/2010ApJ...710..755G} {710, 755}

\bibitem[\protect\citeauthoryear{{Gaspari}, {Ruszkowski}  \& {Oh}}{{Gaspari}
  et~al.}{2013}]{Gaspari_2013}
{Gaspari} M.,  {Ruszkowski} M.,   {Oh} S.~P.,  2013, \mn@doi [\mnras]
  {10.1093/mnras/stt692}, \href
  {http://adsabs.harvard.edu/abs/2013MNRAS.432.3401G} {432, 3401}

\bibitem[\protect\citeauthoryear{{Gebhardt}, {Adams}, {Richstone}, {Lauer},
  {Faber}, {G{\"u}ltekin}, {Murphy}  \& {Tremaine}}{{Gebhardt}
  et~al.}{2011}]{Gebhardt_2011}
{Gebhardt} K.,  {Adams} J.,  {Richstone} D.,  {Lauer} T.~R.,  {Faber} S.~M.,
  {G{\"u}ltekin} K.,  {Murphy} J.,   {Tremaine} S.,  2011, \mn@doi [\apj]
  {10.1088/0004-637X/729/2/119}, \href
  {http://adsabs.harvard.edu/abs/2011ApJ...729..119G} {729, 119}

\bibitem[\protect\citeauthoryear{{Ghez} et~al.,}{{Ghez}
  et~al.}{2003}]{Ghez_2003}
{Ghez} A.~M.,  et~al., 2003, \mn@doi [\apjl] {10.1086/374804}, \href
  {http://adsabs.harvard.edu/abs/2003ApJ...586L.127G} {586, L127}

\bibitem[\protect\citeauthoryear{{Hawley}, {Balbus}  \& {Stone}}{{Hawley}
  et~al.}{2001}]{Hawley_2001}
{Hawley} J.~F.,  {Balbus} S.~A.,   {Stone} J.~M.,  2001, \mn@doi [\apjl]
  {10.1086/320931}, \href {http://adsabs.harvard.edu/abs/2001ApJ...554L..49H}
  {554, L49}

\bibitem[\protect\citeauthoryear{{Ho}}{{Ho}}{2008}]{Ho_2008}
{Ho} L.~C.,  2008, \mn@doi [\araa] {10.1146/annurev.astro.45.051806.110546},
  \href {http://adsabs.harvard.edu/abs/2008ARA%26A..46..475H} {46, 475}

\bibitem[\protect\citeauthoryear{{Ho}}{{Ho}}{2009}]{Ho_2009}
{Ho} L.~C.,  2009, \mn@doi [\apj] {10.1088/0004-637X/699/1/626}, \href
  {http://adsabs.harvard.edu/abs/2009ApJ...699..626H} {699, 626}

\bibitem[\protect\citeauthoryear{{Hopkins}, {Richards}  \&
  {Hernquist}}{{Hopkins} et~al.}{2007}]{Hopkins_2007}
{Hopkins} P.~F.,  {Richards} G.~T.,   {Hernquist} L.,  2007, \mn@doi [\apj]
  {10.1086/509629}, \href {http://adsabs.harvard.edu/abs/2007ApJ...654..731H}
  {654, 731}

\bibitem[\protect\citeauthoryear{{Ichimaru}}{{Ichimaru}}{1977}]{Ichimaru_1977}
{Ichimaru} S.,  1977, \mn@doi [\apj] {10.1086/155314}, \href
  {http://adsabs.harvard.edu/abs/1977ApJ...214..840I} {214, 840}

\bibitem[\protect\citeauthoryear{{Igumenshchev} \& {Abramowicz}}{{Igumenshchev}
  \& {Abramowicz}}{1999}]{IA_1999}
{Igumenshchev} I.~V.,  {Abramowicz} M.~A.,  1999, \mn@doi [\mnras]
  {10.1046/j.1365-8711.1999.02220.x}, \href
  {http://adsabs.harvard.edu/abs/1999MNRAS.303..309I} {303, 309}

\bibitem[\protect\citeauthoryear{{Igumenshchev} \& {Abramowicz}}{{Igumenshchev}
  \& {Abramowicz}}{2000}]{IA_2000}
{Igumenshchev} I.~V.,  {Abramowicz} M.~A.,  2000, \mn@doi [\apjs]
  {10.1086/317354}, \href {http://adsabs.harvard.edu/abs/2000ApJS..130..463I}
  {130, 463}

\bibitem[\protect\citeauthoryear{{Igumenshchev}, {Abramowicz}  \&
  {Narayan}}{{Igumenshchev} et~al.}{2000}]{IAN_2000}
{Igumenshchev} I.~V.,  {Abramowicz} M.~A.,   {Narayan} R.,  2000, \mn@doi
  [\apjl] {10.1086/312755}, \href
  {http://cdsads.u-strasbg.fr/abs/2000ApJ...537L..27I} {537, L27}

\bibitem[\protect\citeauthoryear{{Igumenshchev}, {Narayan}  \&
  {Abramowicz}}{{Igumenshchev} et~al.}{2003}]{INA_2003}
{Igumenshchev} I.~V.,  {Narayan} R.,   {Abramowicz} M.~A.,  2003, \mn@doi
  [\apj] {10.1086/375769}, \href
  {http://adsabs.harvard.edu/abs/2003ApJ...592.1042I} {592, 1042}

\bibitem[\protect\citeauthoryear{{Inayoshi}, {Haiman}  \&
  {Ostriker}}{{Inayoshi} et~al.}{2016}]{IHO_2016}
{Inayoshi} K.,  {Haiman} Z.,   {Ostriker} J.~P.,  2016, \mn@doi [\mnras]
  {10.1093/mnras/stw836}, \href
  {http://adsabs.harvard.edu/abs/2016MNRAS.459.3738I} {459, 3738}

\bibitem[\protect\citeauthoryear{{Ioka}, {Matsumoto}, {Teraki}, {Kashiyama}  \&
  {Murase}}{{Ioka} et~al.}{2017}]{Ioka_2017}
{Ioka} K.,  {Matsumoto} T.,  {Teraki} Y.,  {Kashiyama} K.,   {Murase} K.,
  2017, \mn@doi [\mnras] {10.1093/mnras/stx1337}, \href
  {http://adsabs.harvard.edu/abs/2017MNRAS.470.3332I} {470, 3332}

\bibitem[\protect\citeauthoryear{{Johnson} \& {Quataert}}{{Johnson} \&
  {Quataert}}{2007}]{Johnson_Quataert_2007}
{Johnson} B.~M.,  {Quataert} E.,  2007, \mn@doi [\apj] {10.1086/513065}, \href
  {http://cdsads.u-strasbg.fr/abs/2007ApJ...660.1273J} {660, 1273}

\bibitem[\protect\citeauthoryear{{Kato}, {Fukue}  \& {Mineshige}}{{Kato}
  et~al.}{2008}]{Kato_2008}
{Kato} S.,  {Fukue} J.,   {Mineshige} S.,  2008, {Black-Hole Accretion Disks
  --- Towards a New Paradigm ---}

\bibitem[\protect\citeauthoryear{{Kuiper}, {Klahr}, {Beuther}  \&
  {Henning}}{{Kuiper} et~al.}{2010}]{Kuiper_2010}
{Kuiper} R.,  {Klahr} H.,  {Beuther} H.,   {Henning} T.,  2010, \mn@doi [\apj]
  {10.1088/0004-637X/722/2/1556}, \href
  {http://adsabs.harvard.edu/abs/2010ApJ...722.1556K} {722, 1556}

\bibitem[\protect\citeauthoryear{{Kuiper}, {Klahr}, {Beuther}  \&
  {Henning}}{{Kuiper} et~al.}{2011}]{Kuiper_2011}
{Kuiper} R.,  {Klahr} H.,  {Beuther} H.,   {Henning} T.,  2011, \mn@doi [\apj]
  {10.1088/0004-637X/732/1/20}, \href
  {http://adsabs.harvard.edu/abs/2011ApJ...732...20K} {732, 20}

\bibitem[\protect\citeauthoryear{{Kuo} et~al.,}{{Kuo} et~al.}{2014}]{Kuo_2014}
{Kuo} C.~Y.,  et~al., 2014, \mn@doi [\apjl] {10.1088/2041-8205/783/2/L33},
  \href {http://adsabs.harvard.edu/abs/2014ApJ...783L..33K} {783, L33}

\bibitem[\protect\citeauthoryear{{Li}, {Ostriker}  \& {Sunyaev}}{{Li}
  et~al.}{2013}]{LOS_2013}
{Li} J.,  {Ostriker} J.,   {Sunyaev} R.,  2013, \mn@doi [\apj]
  {10.1088/0004-637X/767/2/105}, \href
  {http://adsabs.harvard.edu/abs/2013ApJ...767..105L} {767, 105}

\bibitem[\protect\citeauthoryear{{Machida}, {Matsumoto}  \&
  {Mineshige}}{{Machida} et~al.}{2001}]{Machida_2001}
{Machida} M.,  {Matsumoto} R.,   {Mineshige} S.,  2001, \mn@doi [\pasj]
  {10.1093/pasj/53.1.L1}, \href
  {http://adsabs.harvard.edu/abs/2001PASJ...53L...1M} {53, L1}

\bibitem[\protect\citeauthoryear{{Maron}, {Chandran}  \& {Blackman}}{{Maron}
  et~al.}{2004}]{Maron_2004}
{Maron} J.,  {Chandran} B.~D.,   {Blackman} E.,  2004, \mn@doi [Physical Review
  Letters] {10.1103/PhysRevLett.92.045001}, \href
  {http://adsabs.harvard.edu/abs/2004PhRvL..92d5001M} {92, 045001}

\bibitem[\protect\citeauthoryear{{Matsumoto} \& {Tajima}}{{Matsumoto} \&
  {Tajima}}{1995}]{Matsumoto_1995}
{Matsumoto} R.,  {Tajima} T.,  1995, \mn@doi [\apj] {10.1086/175739}, \href
  {http://adsabs.harvard.edu/abs/1995ApJ...445..767M} {445, 767}

\bibitem[\protect\citeauthoryear{{Matsumoto}, {Teraki}  \& {Ioka}}{{Matsumoto}
  et~al.}{2017}]{Matsumoto_2017}
{Matsumoto} T.,  {Teraki} Y.,   {Ioka} K.,  2017, preprint, \href
  {http://adsabs.harvard.edu/abs/2017arXiv170405047M} {} (\mn@eprint {arXiv}
  {1704.05047})

\bibitem[\protect\citeauthoryear{{McCourt}, {Parrish}, {Sharma}  \&
  {Quataert}}{{McCourt} et~al.}{2011}]{McCourt_2011}
{McCourt} M.,  {Parrish} I.~J.,  {Sharma} P.,   {Quataert} E.,  2011, \mn@doi
  [\mnras] {10.1111/j.1365-2966.2011.18216.x}, \href
  {http://adsabs.harvard.edu/abs/2011MNRAS.413.1295M} {413, 1295}

\bibitem[\protect\citeauthoryear{{McKinney} \& {Gammie}}{{McKinney} \&
  {Gammie}}{2004}]{McKinney_&_Gammie_2004}
{McKinney} J.~C.,  {Gammie} C.~F.,  2004, \mn@doi [\apj] {10.1086/422244},
  \href {http://cdsads.u-strasbg.fr/abs/2004ApJ...611..977M} {611, 977}

\bibitem[\protect\citeauthoryear{{Mignone}, {Bodo}, {Massaglia}, {Matsakos},
  {Tesileanu}, {Zanni}  \& {Ferrari}}{{Mignone} et~al.}{2007}]{Mignone_2007}
{Mignone} A.,  {Bodo} G.,  {Massaglia} S.,  {Matsakos} T.,  {Tesileanu} O.,
  {Zanni} C.,   {Ferrari} A.,  2007, \mn@doi [\apjs] {10.1086/513316}, \href
  {http://adsabs.harvard.edu/abs/2007ApJS..170..228M} {170, 228}

\bibitem[\protect\citeauthoryear{{Milosavljevi{\'c}}, {Bromm}, {Couch}  \&
  {Oh}}{{Milosavljevi{\'c}} et~al.}{2009}]{Milosavljevic_2009}
{Milosavljevi{\'c}} M.,  {Bromm} V.,  {Couch} S.~M.,   {Oh} S.~P.,  2009,
  \mn@doi [\apj] {10.1088/0004-637X/698/1/766}, \href
  {http://adsabs.harvard.edu/abs/2009ApJ...698..766M} {698, 766}

\bibitem[\protect\citeauthoryear{{Najarro}, {Krabbe}, {Genzel}, {Lutz},
  {Kudritzki}  \& {Hillier}}{{Najarro} et~al.}{1997}]{Najarro_1997}
{Najarro} F.,  {Krabbe} A.,  {Genzel} R.,  {Lutz} D.,  {Kudritzki} R.~P.,
  {Hillier} D.~J.,  1997, \aap, \href
  {http://adsabs.harvard.edu/abs/1997A%26A...325..700N} {325, 700}

\bibitem[\protect\citeauthoryear{{Narayan} \& {Fabian}}{{Narayan} \&
  {Fabian}}{2011}]{Narayan_Fabian_2011}
{Narayan} R.,  {Fabian} A.~C.,  2011, \mn@doi [\mnras]
  {10.1111/j.1365-2966.2011.18987.x}, \href
  {http://adsabs.harvard.edu/abs/2011MNRAS.415.3721N} {415, 3721}

\bibitem[\protect\citeauthoryear{{Narayan} \& {Medvedev}}{{Narayan} \&
  {Medvedev}}{2001}]{Narayan_Medvedev_2001}
{Narayan} R.,  {Medvedev} M.~V.,  2001, \mn@doi [\apjl] {10.1086/338325}, \href
  {http://adsabs.harvard.edu/abs/2001ApJ...562L.129N} {562, L129}

\bibitem[\protect\citeauthoryear{{Narayan} \& {Yi}}{{Narayan} \&
  {Yi}}{1994}]{NY_1994}
{Narayan} R.,  {Yi} I.,  1994, \mn@doi [\apjl] {10.1086/187381}, \href
  {http://adsabs.harvard.edu/abs/1994ApJ...428L..13N} {428, L13}

\bibitem[\protect\citeauthoryear{{Narayan} \& {Yi}}{{Narayan} \&
  {Yi}}{1995}]{NY_1995}
{Narayan} R.,  {Yi} I.,  1995, \mn@doi [\apj] {10.1086/175599}, \href
  {http://adsabs.harvard.edu/abs/1995ApJ...444..231N} {444, 231}

\bibitem[\protect\citeauthoryear{{Narayan}, {Igumenshchev}  \&
  {Abramowicz}}{{Narayan} et~al.}{2000}]{Narayan_2000}
{Narayan} R.,  {Igumenshchev} I.~V.,   {Abramowicz} M.~A.,  2000, \mn@doi
  [\apj] {10.1086/309268}, \href
  {http://adsabs.harvard.edu/abs/2000ApJ...539..798N} {539, 798}

\bibitem[\protect\citeauthoryear{{Narayan}, {S\c{a}dowski}, {Penna}  \&
  {Kulkarni}}{{Narayan} et~al.}{2012}]{Narayan_2012}
{Narayan} R.,  {S\c{a}dowski} A.,  {Penna} R.~F.,   {Kulkarni} A.~K.,  2012,
  \mn@doi [\mnras] {10.1111/j.1365-2966.2012.22002.x}, \href
  {http://adsabs.harvard.edu/abs/2012MNRAS.426.3241N} {426, 3241}

\bibitem[\protect\citeauthoryear{{Ohsuga}, {Mineshige}, {Mori}  \&
  {Kato}}{{Ohsuga} et~al.}{2009}]{Ohsuga_2009}
{Ohsuga} K.,  {Mineshige} S.,  {Mori} M.,   {Kato} Y.,  2009, \mn@doi [\pasj]
  {10.1093/pasj/61.3.L7}, \href
  {http://adsabs.harvard.edu/abs/2009PASJ...61L...7O} {61, L7}

\bibitem[\protect\citeauthoryear{{Ostriker}, {Weaver}, {Yahil}  \&
  {McCray}}{{Ostriker} et~al.}{1976}]{Ostriker_1976}
{Ostriker} J.~P.,  {Weaver} R.,  {Yahil} A.,   {McCray} R.,  1976, \mn@doi
  [\apjl] {10.1086/182233}, \href
  {http://adsabs.harvard.edu/abs/1976ApJ...208L..61O} {208, L61}

\bibitem[\protect\citeauthoryear{{Papaloizou} \& {Pringle}}{{Papaloizou} \&
  {Pringle}}{1984}]{Papaloizou_Pringle_1984}
{Papaloizou} J.~C.~B.,  {Pringle} J.~E.,  1984, \mn@doi [\mnras]
  {10.1093/mnras/208.4.721}, \href
  {http://adsabs.harvard.edu/abs/1984MNRAS.208..721P} {208, 721}

\bibitem[\protect\citeauthoryear{{Park}}{{Park}}{1990}]{Park_1990a}
{Park} M.-G.,  1990, \mn@doi [\apj] {10.1086/168668}, \href
  {http://esoads.eso.org/abs/1990ApJ...354...64P} {354, 64}

\bibitem[\protect\citeauthoryear{{Park} \& {Ricotti}}{{Park} \&
  {Ricotti}}{2011}]{PR_2011}
{Park} K.,  {Ricotti} M.,  2011, \mn@doi [\apj] {10.1088/0004-637X/739/1/2},
  \href {http://adsabs.harvard.edu/abs/2011ApJ...739....2P} {739, 2}

\bibitem[\protect\citeauthoryear{{Perna}, {Narayan}, {Rybicki}, {Stella}  \&
  {Treves}}{{Perna} et~al.}{2003}]{Perna_2003}
{Perna} R.,  {Narayan} R.,  {Rybicki} G.,  {Stella} L.,   {Treves} A.,  2003,
  \mn@doi [\apj] {10.1086/377091}, \href
  {http://cdsads.u-strasbg.fr/abs/2003ApJ...594..936P} {594, 936}

\bibitem[\protect\citeauthoryear{{Pringle}}{{Pringle}}{1981}]{Pringle_1981}
{Pringle} J.~E.,  1981, \mn@doi [\araa] {10.1146/annurev.aa.19.090181.001033},
  \href {http://adsabs.harvard.edu/abs/1981ARA%26A..19..137P} {19, 137}

\bibitem[\protect\citeauthoryear{{Proga}}{{Proga}}{2007}]{Proga_2007}
{Proga} D.,  2007, \mn@doi [\apj] {10.1086/515389}, \href
  {http://adsabs.harvard.edu/abs/2007ApJ...661..693P} {661, 693}

\bibitem[\protect\citeauthoryear{{Proga} \& {Begelman}}{{Proga} \&
  {Begelman}}{2003}]{Proga_2003b}
{Proga} D.,  {Begelman} M.~C.,  2003, \mn@doi [\apj] {10.1086/375773}, \href
  {http://adsabs.harvard.edu/abs/2003ApJ...592..767P} {592, 767}

\bibitem[\protect\citeauthoryear{{Quataert}}{{Quataert}}{2004}]{Quataert_2004}
{Quataert} E.,  2004, \mn@doi [\apj] {10.1086/422973}, \href
  {http://adsabs.harvard.edu/abs/2004ApJ...613..322Q} {613, 322}

\bibitem[\protect\citeauthoryear{{Quataert}}{{Quataert}}{2008}]{Quataert_2008}
{Quataert} E.,  2008, \mn@doi [\apj] {10.1086/525248}, \href
  {http://adsabs.harvard.edu/abs/2008ApJ...673..758Q} {673, 758}

\bibitem[\protect\citeauthoryear{{Quataert} \& {Gruzinov}}{{Quataert} \&
  {Gruzinov}}{2000}]{Quataert_2000}
{Quataert} E.,  {Gruzinov} A.,  2000, \mn@doi [\apj] {10.1086/309267}, \href
  {http://adsabs.harvard.edu/abs/2000ApJ...539..809Q} {539, 809}

\bibitem[\protect\citeauthoryear{{Roberts}, {Jiang}, {Wang}  \&
  {Ostriker}}{{Roberts} et~al.}{2017}]{Roberts_2017}
{Roberts} S.~R.,  {Jiang} Y.-F.,  {Wang} Q.~D.,   {Ostriker} J.~P.,  2017,
  \mn@doi [\mnras] {10.1093/mnras/stw2995}, \href
  {http://adsabs.harvard.edu/abs/2017MNRAS.466.1477R} {466, 1477}

\bibitem[\protect\citeauthoryear{{Russell}, {Fabian}, {McNamara}  \&
  {Broderick}}{{Russell} et~al.}{2015}]{Russell_2015}
{Russell} H.~R.,  {Fabian} A.~C.,  {McNamara} B.~R.,   {Broderick} A.~E.,
  2015, \mn@doi [\mnras] {10.1093/mnras/stv954}, \href
  {http://adsabs.harvard.edu/abs/2015MNRAS.451..588R} {451, 588}

\bibitem[\protect\citeauthoryear{{Ryan}, {Ressler}, {Dolence}, {Tchekhovskoy},
  {Gammie}  \& {Quataert}}{{Ryan} et~al.}{2017}]{Ryan_2017}
{Ryan} B.~R.,  {Ressler} S.~M.,  {Dolence} J.~C.,  {Tchekhovskoy} A.,  {Gammie}
  C.,   {Quataert} E.,  2017, \mn@doi [\apjl] {10.3847/2041-8213/aa8034}, \href
  {http://adsabs.harvard.edu/abs/2017ApJ...844L..24R} {844, L24}

\bibitem[\protect\citeauthoryear{{Sano}, {Inutsuka}, {Turner}  \&
  {Stone}}{{Sano} et~al.}{2004}]{Sano_2004}
{Sano} T.,  {Inutsuka} S.-i.,  {Turner} N.~J.,   {Stone} J.~M.,  2004, \mn@doi
  [\apj] {10.1086/382184}, \href
  {http://adsabs.harvard.edu/abs/2004ApJ...605..321S} {605, 321}

\bibitem[\protect\citeauthoryear{{S{\c a}dowski}, {Wielgus}, {Narayan},
  {Abarca}, {McKinney}  \& {Chael}}{{S{\c a}dowski}
  et~al.}{2017}]{Sadowski_2017}
{S{\c a}dowski} A.,  {Wielgus} M.,  {Narayan} R.,  {Abarca} D.,  {McKinney}
  J.~C.,   {Chael} A.,  2017, \mn@doi [\mnras] {10.1093/mnras/stw3116}, \href
  {http://adsabs.harvard.edu/abs/2017MNRAS.466..705S} {466, 705}

\bibitem[\protect\citeauthoryear{{Shakura} \& {Sunyaev}}{{Shakura} \&
  {Sunyaev}}{1973}]{SS_1973}
{Shakura} N.~I.,  {Sunyaev} R.~A.,  1973, \aap, \href
  {http://adsabs.harvard.edu/abs/1973A%26A....24..337S} {24, 337}

\bibitem[\protect\citeauthoryear{{Shapiro}}{{Shapiro}}{1973}]{Shapiro_1973}
{Shapiro} S.~L.,  1973, \mn@doi [\apj] {10.1086/151982}, \href
  {http://esoads.eso.org/abs/1973ApJ...180..531S} {180, 531}

\bibitem[\protect\citeauthoryear{{Sharma}, {Quataert}  \& {Stone}}{{Sharma}
  et~al.}{2008}]{Sharma_2008}
{Sharma} P.,  {Quataert} E.,   {Stone} J.~M.,  2008, \mn@doi [\mnras]
  {10.1111/j.1365-2966.2008.13686.x}, \href
  {http://adsabs.harvard.edu/abs/2008MNRAS.389.1815S} {389, 1815}

\bibitem[\protect\citeauthoryear{{Shcherbakov} \& {Baganoff}}{{Shcherbakov} \&
  {Baganoff}}{2010}]{Shcherbakov_2010}
{Shcherbakov} R.~V.,  {Baganoff} F.~K.,  2010, \mn@doi [\apj]
  {10.1088/0004-637X/716/1/504}, \href
  {http://cdsads.u-strasbg.fr/abs/2010ApJ...716..504S} {716, 504}

\bibitem[\protect\citeauthoryear{{Spitzer}}{{Spitzer}}{1962}]{Spitzer_1962}
{Spitzer} L.,  1962, {Physics of Fully Ionized Gases}

\bibitem[\protect\citeauthoryear{{Stone} \& {Norman}}{{Stone} \&
  {Norman}}{1992}]{Stone_Norman_1992}
{Stone} J.~M.,  {Norman} M.~L.,  1992, \mn@doi [\apjs] {10.1086/191680}, \href
  {http://adsabs.harvard.edu/abs/1992ApJS...80..753S} {80, 753}

\bibitem[\protect\citeauthoryear{{Stone} \& {Pringle}}{{Stone} \&
  {Pringle}}{2001}]{Stone_Pringle_2001}
{Stone} J.~M.,  {Pringle} J.~E.,  2001, \mn@doi [\mnras]
  {10.1046/j.1365-8711.2001.04138.x}, \href
  {http://adsabs.harvard.edu/abs/2001MNRAS.322..461S} {322, 461}

\bibitem[\protect\citeauthoryear{{Stone}, {Hawley}, {Gammie}  \&
  {Balbus}}{{Stone} et~al.}{1996}]{Stone_1996}
{Stone} J.~M.,  {Hawley} J.~F.,  {Gammie} C.~F.,   {Balbus} S.~A.,  1996,
  \mn@doi [\apj] {10.1086/177280}, \href
  {http://adsabs.harvard.edu/abs/1996ApJ...463..656S} {463, 656}

\bibitem[\protect\citeauthoryear{{Stone}, {Pringle}  \& {Begelman}}{{Stone}
  et~al.}{1999}]{Stone_1999}
{Stone} J.~M.,  {Pringle} J.~E.,   {Begelman} M.~C.,  1999, \mn@doi [\mnras]
  {10.1046/j.1365-8711.1999.03024.x}, \href
  {http://adsabs.harvard.edu/abs/1999MNRAS.310.1002S} {310, 1002}

\bibitem[\protect\citeauthoryear{{Walsh}, {Barth}, {Ho}  \& {Sarzi}}{{Walsh}
  et~al.}{2013}]{Walsh_2013}
{Walsh} J.~L.,  {Barth} A.~J.,  {Ho} L.~C.,   {Sarzi} M.,  2013, \mn@doi [\apj]
  {10.1088/0004-637X/770/2/86}, \href
  {http://adsabs.harvard.edu/abs/2013ApJ...770...86W} {770, 86}

\bibitem[\protect\citeauthoryear{{Yuan}, {Quataert}  \& {Narayan}}{{Yuan}
  et~al.}{2003}]{Yuan_2003}
{Yuan} F.,  {Quataert} E.,   {Narayan} R.,  2003, \mn@doi [\apj]
  {10.1086/378716}, \href {http://adsabs.harvard.edu/abs/2003ApJ...598..301Y}
  {598, 301}

\bibitem[\protect\citeauthoryear{{Yuan}, {Wu}  \& {Bu}}{{Yuan}
  et~al.}{2012}]{Yuan_2012a}
{Yuan} F.,  {Wu} M.,   {Bu} D.,  2012, \mn@doi [\apj]
  {10.1088/0004-637X/761/2/129}, \href
  {http://cdsads.u-strasbg.fr/abs/2012ApJ...761..129Y} {761, 129}

\makeatother
\end{thebibliography}
}


\appendix
\section{Numerical tests}
\label{sec:app}

In this appendix, we discuss the effects of numerical resolutions,
outer boundary conditions, initial conditions, prescriptions of viscosity on our results.
As a reference, we define our fiducial case with $R_{\rm C}/R_{\rm B}=0.1$ 
and $\alpha=0.01$, where (1) the number of grid cells is 
$N_r \times N_\theta =512 \times 256$, (2) the initial condition is given by
a rotating equilibrium density configuration by Eq. (\ref{eq:torus}), 
and (3) the additional viscosity which would be driven by rotational instability is assumed
(see Eq. \ref{eq:alpha_j}).

\begin{figure}
\begin{center}
\includegraphics[width=82mm]{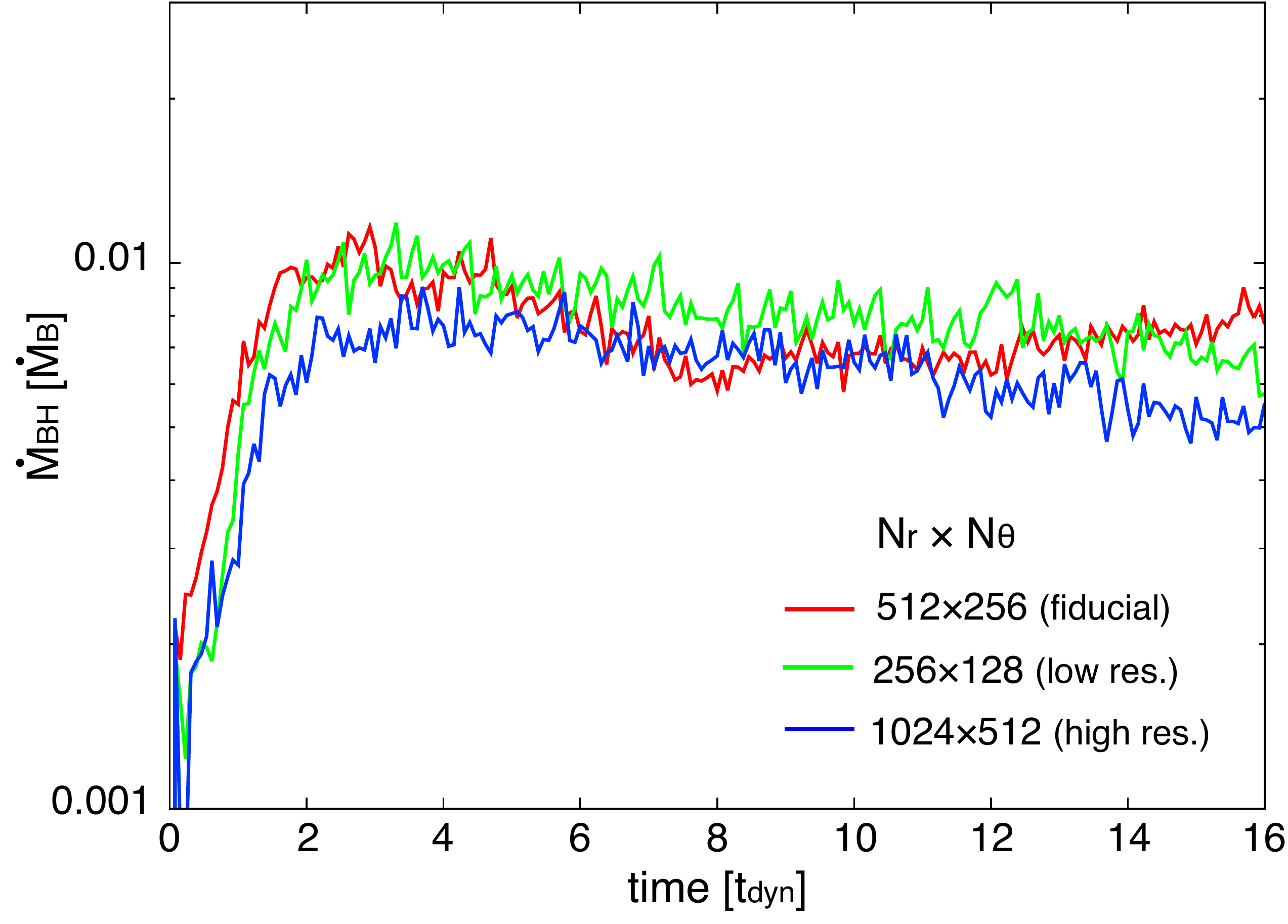}
\caption{Time evolution of the net accretion onto a sink cell (in units of the Bondi rate)
for $R_{\rm C}/R_{\rm B}=0.1$ and $\alpha=0.01$.
Simulation results with three different resolutions are shown:
$N_r \times N_\theta =512 \times 256$ (fiducial, red), $256 \times 256$ (low resolution, green)
and $1024 \times 512$ (high resolution, blue).
}
\label{fig:res}
\end{center}
\end{figure}

Fig. \ref{fig:res} shows the time evolution of the net accretion rates onto the BH (i.e., a sink cell at the center)
with different numerical resolutions: $N_r \times N_\theta =512 \times 256$ (fiducial, red), 
$256 \times 256$ (low resolution, green) and $1024 \times 512$ (high resolution, blue).
For the three cases, the time-averaged accretion rate is 
$\dot{M}/\dot{M}_{\rm B}=7.1\times 10^{-3}$ (red), $7.9\times 10^{-3}$ (green), 
and $6.3\times 10^{-3}$ (blue), respectively.
The estimated errors are at most $\sim 20~\%$.

\begin{figure}
\begin{center}
\includegraphics[width=82mm]{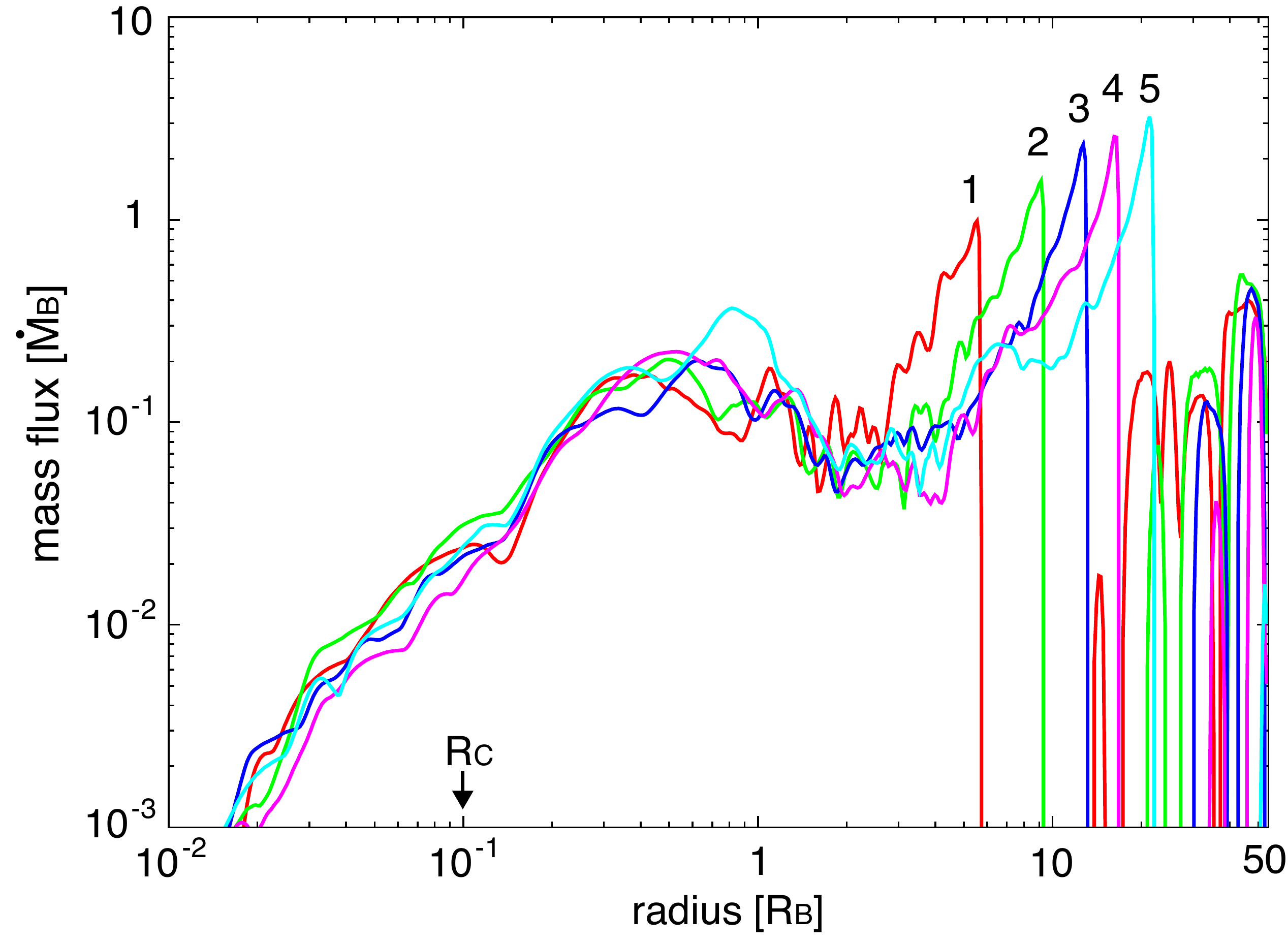}
\caption{Time evolution of radial profiles of the outflow rates
for $R_{\rm C}/R_{\rm B}=0.1$ and $\alpha=0.01$.
Different curves show the time evolution: $t/t_{\rm dyn}=6$ (1, red), $10$ (2, green), $14$ (3, blue), 
$18$ (4, magenta) and $23$ (5, light blue).
Although the outflow propagates, accumulating the ambient gas, the outflow rate inside the Bondi radius 
is steady. This result ensures the stationarity of the accretion flow in the inner region.
}
\label{fig:outflow}
\end{center}
\end{figure}

As seen in Figs.~\ref{fig:r_Mdot} and \ref{fig:rho_temp_ad}, a significant fraction of the gas is
outflowing towards the polar directions and the rate dominates the inflow rate outside the Bondi radius.
In order to ensure stationarity of the accretion system, we study the time evolution of the outflow rate.
Fig.~\ref{fig:outflow} shows radial profiles of the outflow rates (i.e., $v_r>0$, see Eq. \ref{eq:Mout})
for different elapsed times ($6\leq t/t_{\rm dyn}\leq 23$).
This component propagates outwards, accumulating the ambient gas which has a uniform distribution
and no accretion initially at $r>R_{\rm B}$.
However, the outflow rates inside the Bondi radius seems steady, and the rates even outside the Bondi radius 
have nearly converged ($r\la 7~R_{\rm B}$).
Therefore, our long-term simulations allow us to ensure the stationarity of the accretion flow in the inner region.

\begin{figure}
\begin{center}
\includegraphics[width=82mm]{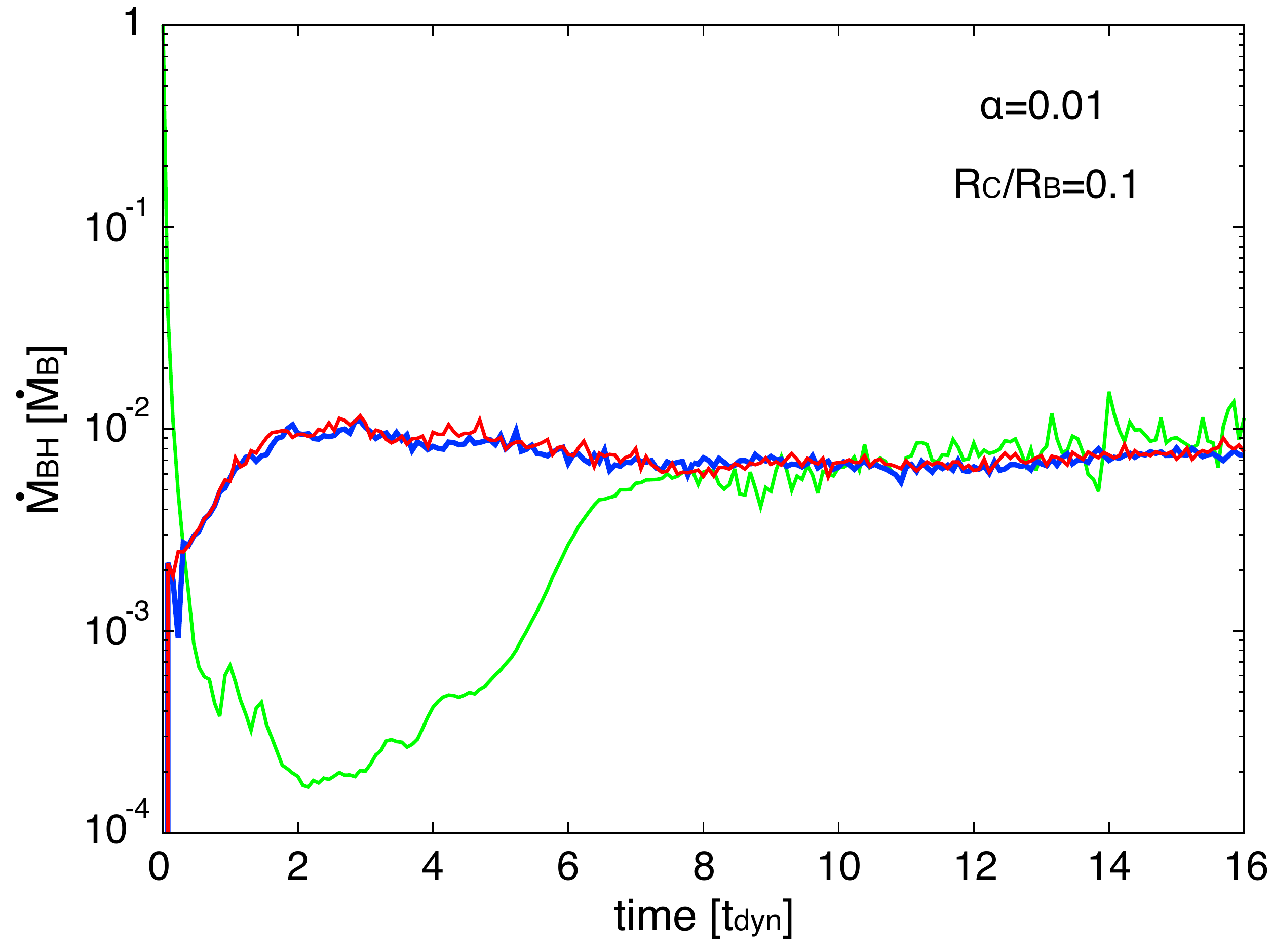}
\caption{Time evolution of the net accretion onto a sink cell (in units of the Bondi rate)
for $R_{\rm C}/R_{\rm B}=0.1$ and $\alpha=0.01$.
Red curve presents the fiducial case as shown in Fig. \ref{fig:t_Mdot_j}.
Green one is the result where the simulation starts from the Bondi solution
with a constant angular momentum $j_\infty$.
Blue one is the result where the second term in Eq.(\ref{eq:alpha_j}), which provides
additional viscosity led by rotational instability, is neglected.
}
\label{fig:app_1}
\end{center}
\end{figure}

Fig. \ref{fig:app_1} shows the effect of initial conditions on the time evolution of the accretion rate.
Red and green curve present the fiducial case and the result with a Bondi accretion solution 
added a constant angular momentum $j_\infty$ as the initial conditions.
In the later case, the accretion rate decreases rapidly by several orders of magnitude in the early stage
due to the centrifugal force, begins to increase gradually at $t\ga 2~t_{\rm dyn}$ and 
approaches that in the fiducial case at $t\ga 6~t_{\rm dyn}$.
Therefore, this result clearly shows that the net accretion rate does not depend 
on the choice of the initial conditions.

\begin{figure}
\begin{center}
\includegraphics[width=82mm]{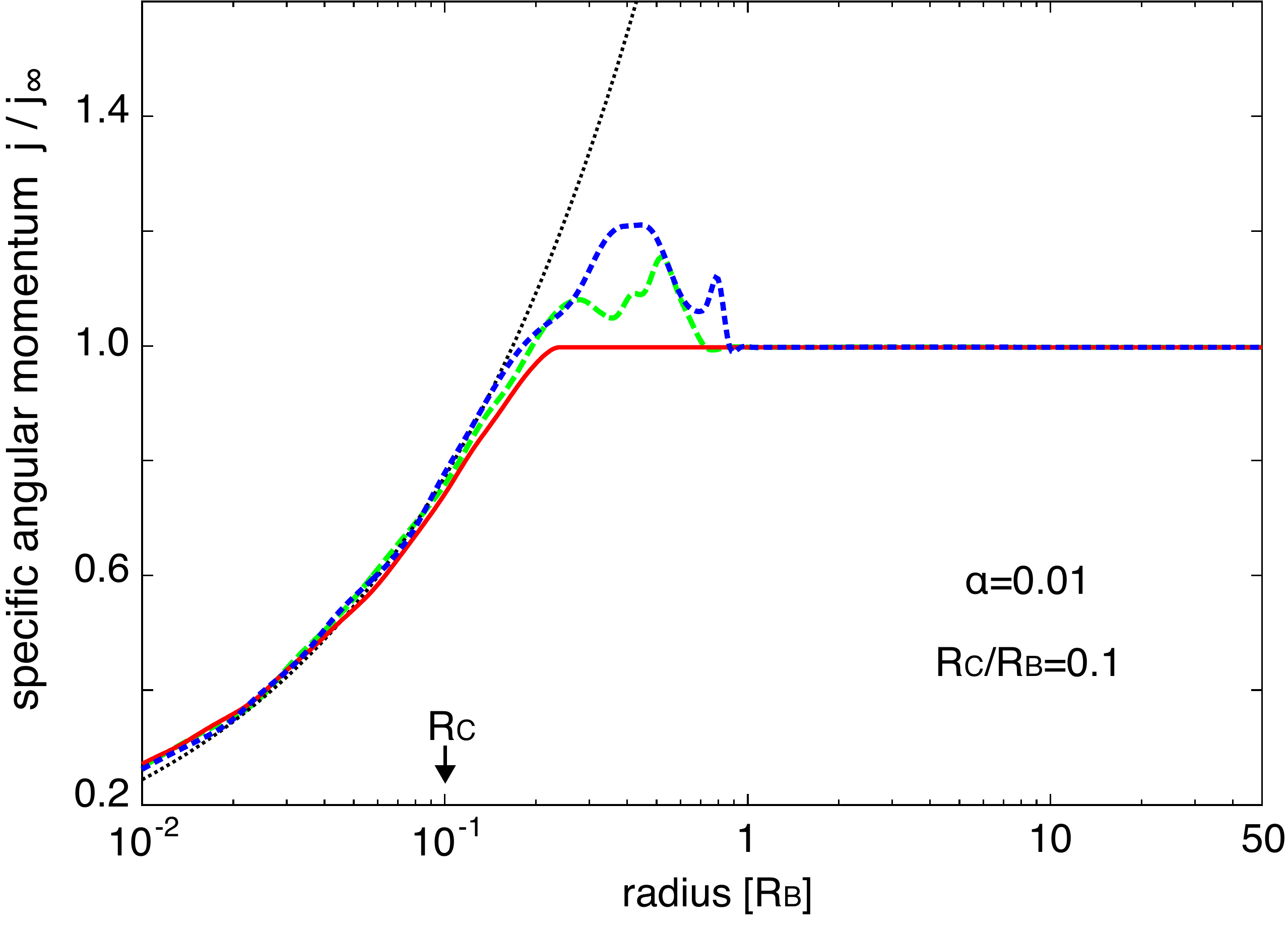}
\caption{Radial profiles of the specific angular momentum normalized by 
$j_\infty (=\sqrt{\beta}R_{\rm B}c_\infty)$
for $R_{\rm C}/R_{\rm B}=0.1$ and $\alpha=0.01$.
Solid (red) curve presents the profile at $t=13~t_{\rm dyn}$ in our fiducial case, 
where the second term in Eq.(\ref{eq:alpha_j}) is included to provide additional viscosity led by rotational instability.
Dashed curves show the results at $t=13~t_{\rm dyn}$ (green) and $16~t_{\rm dyn}$ (blue), respectively,
in cases where the additional viscosity is neglected.
Dotted (black) curve is that expected in a CDAF solution, $j=g(\gamma)\sqrt{GM_\bullet r}$.
}
\label{fig:app_2}
\end{center}
\end{figure}

Finally, we study the effect of the additional viscosity given by
the second term in Eq.(\ref{eq:alpha_j}), which would be led by rotational instability.
Blue curve in Fig. \ref{fig:app_1} shows the accretion rate without the additional term, and 
almost identical to that in the fiducial case.
Fig. \ref{fig:app_2} shows radial profiles of the specific angular momentum normalized by 
$j_\infty (=\sqrt{\beta}R_{\rm B}c_\infty)$ for $R_{\rm C}/R_{\rm B}=0.1$ and $\alpha=0.01$.
Solid (red) curve presents the result at $t=13~t_{\rm dyn}$ in our fiducial case.
The profile has a Keplerian-like profile expected from a CDAF solution inside the centrifugal radius,
$j=g(\gamma)\sqrt{GM_\bullet r}$ (dotted black), and a constant value outside the torus ($r\ga 2~R_{\rm C}$).
Two dashed curves show the cases without additional viscosity in the second term in Eq.(\ref{eq:alpha_j})
at two different elapsed time of $t=13~t_{\rm dyn}$ (green) and $t=16~t_{\rm dyn}$ (blue).
Compared to the fiducial case, the profile at the same elapsed time (green curve) have an excess of 
the angular momentum from the boundary value of $j_\infty$ outside the torus.
Since viscosity does not work outside the torus, the angular momentum is accumulated and the bump 
structure grows with time (see blue curve).
As discussed in \S\ref{sec:method}, the regions with negative gradients of the angular momentum (i.e., $dj/dr<0$)
are unstable and would produce turbulence, which allows additional angular momentum transport.
In order to capture the physics of the instability, we need three-dimensional hydrodynamical simulations.
However, we have confirmed that the treatment of the rotational instability does not affect our results.

\end{document}